\input fcptex.arc

\newcount\citeNumber	
	\citeNumber = 0
\newtoks\bibentry
\def\newCite#1#2{%
\global\advance\citeNumber by 1 \the\citeNumber%
\immediate\write\tmpFile{#1 \the\citeNumber}%
\bibentry={#2}%
\immediate\write\biboutFile{\noexpand\item{\the\citeNumber.} \the\bibentry}%
}
\def\newFigure#1{%
\global\advance\figureNumber by 1 \the\figureNumber%
\immediate\write\tmpFile{#1 \the\figureNumber}%
}
\def\newTable#1{%
\global\advance\tableNumber by 1 \uppercase\expandafter{\romannumeral\the\tableNumber}%
\immediate\write\tmpFile{#1 \uppercase{\romannumeral\the\tableNumber}}%
}
\newcount\eqNumber 	
\eqNumber = 0
\def\newEq#1{%
\global\advance\eqNumber by 1 \the\eqNumber%
\immediate\write\tmpFile{#1 \the\eqNumber}%
}
  
\newread\bibFile
\def\makebib{%
\immediate\closeout\biboutFile
\immediate\openin\bibFile=\jobname.bib
\loop
\ifeof\bibFile\global\notdonefalse 
\else {%
\read\bibFile to\Line%
\hangindent=\parindent
\Line
\global\notdonetrue%
}\fi%
\ifnotdone
\repeat
\closein\bibFile
}   

\def\title#1{\centerline{\seventeenbf#1}}
\def\author#1{\centerline{#1}}
\def\caltech{California Institute of Technology}
\def\lauritsen{Lauritsen Laboratory for High Energy Physics}
\def\caltechAddress{Pasadena, California 91125}
\def\institutionCaltech{\centerline{\caltech}
                        \centerline{\lauritsen}
                        \centerline{\caltechAddress}}
                         
\def\abstract#1{\centerline{{\fourteenpoint{\bf Abstract}}}\smallskip \narrower#1}
  
\newcount\chapterNumber
  \chapterNumber = 0
\def\chapter#1#2{\sectionNumber = 0\global\advance\chapterNumber by 1%
\bigskip{\seventeenpoint\centerline{\bf\the\chapterNumber. #2}}%
\immediate\write\tmpFile{#1 \the\chapterNumber}}%

\newcount\sectionNumber 
  \sectionNumber = 0  
\def\section#1#2{\subsectionNumber = 0\global\advance\sectionNumber by 1%
  \bigskip{\fourteenpoint\centerline{\bf\the\chapterNumber.\the\sectionNumber\ #2}}
  \immediate\write\tmpFile{#1 \the\chapterNumber.\the\sectionNumber}}
  
\newcount\subsectionNumber 
  \subsectionNumber = 0  
\def\subsection#1#2{\global\advance\subsectionNumber by 1%
  \medskip{\centerline{\bf\the\chapterNumber.\the\sectionNumber.\the\subsectionNumber\ #2}}
  \immediate\write\tmpFile{#1 \the\chapterNumber.\the\sectionNumber.\the\subsectionNumber}%
}

\twelvepoint

\title{Testing Consistency of Two Histograms{\twelvepoint\footnote*{This work supported in part by the 
U.S.\ Department of Energy under grant DE-FG02-92-ER40701.}}}
\author{Frank Porter}
\institutionCaltech
\centerline{March 7, 2008}

\bigskip

\abstract{Several approaches to testing the hypothesis that two histograms
are drawn from the same distribution are investigated. We note that 
single-sample continuous distribution tests may be adapted to this
two-sample grouped data situation. The difficulty of
not having a fully-specified null hypothesis is an important consideration
in the general case, and care is required in estimating probabilities with
``toy'' Monte Carlo simulations. The performance of several common tests
is compared; no single test performs best in all situations.}

\chapter{ch:intro}{Introduction}

\smallskip

Sometimes we have two histograms and are faced with the question:
``Are they consistent?'' That is, are our two histograms consistent
with having been sampled from the same parent distribution. For example,
we might have a kinematic distribution in two similar channels that we
think should be consistent, and wish to test this hypothesis.
Each histogram represents a sampling from a multivariate Poisson
distribution. The question is whether the means are bin-by-bin
equal between the two distributions. Or, if we are only interested
in ``shape'', are the means related by the same scale factor for all 
bins? We investigate this question in the context of frequency statistics.

For example, consider Fig.~\refno{example1}. Are the two histograms consistent or
can we conclude that they are drawn from different distributions?


\vbox{
\hbox{\centerline{\includegraphics[width=3.5in]{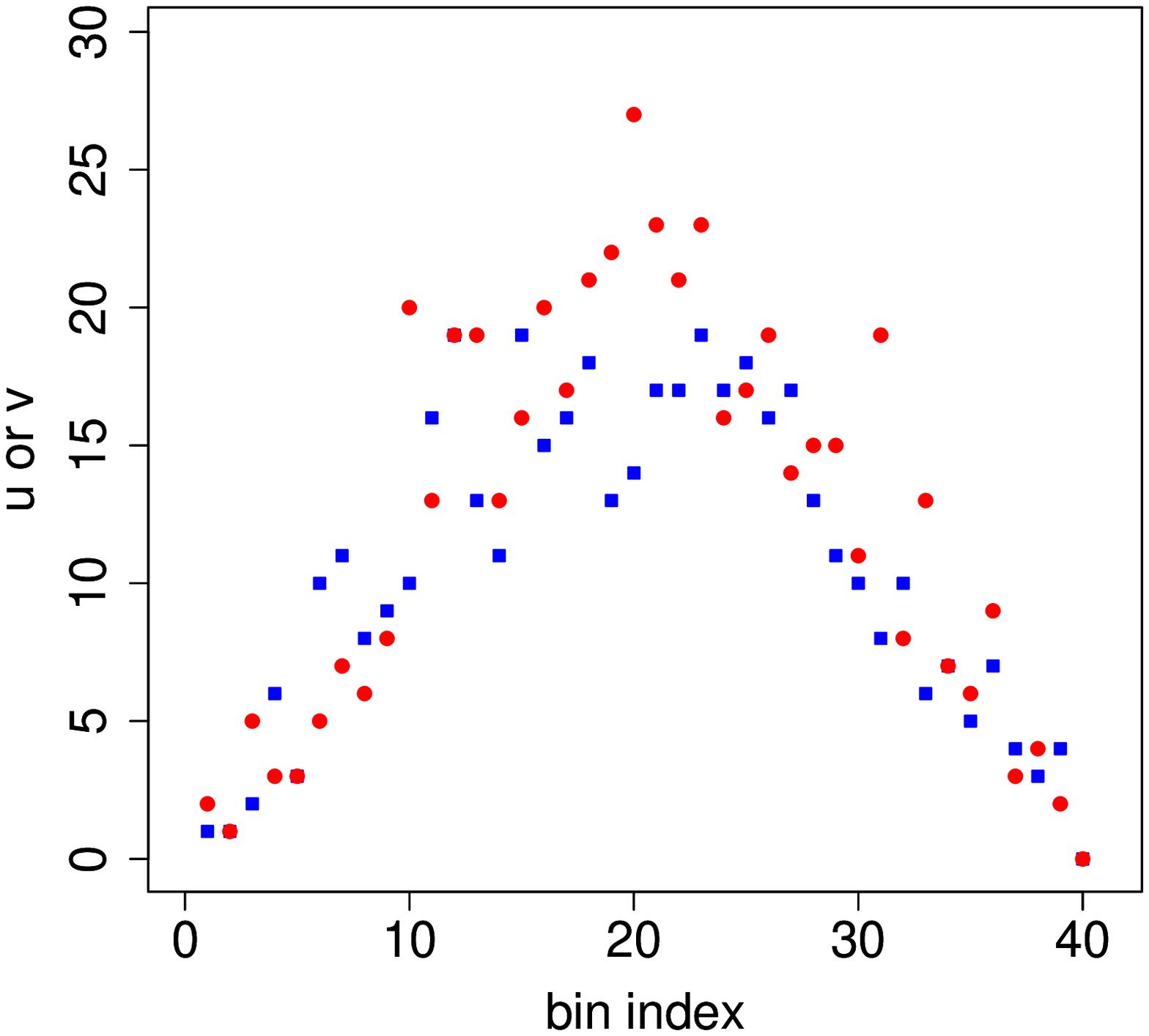}}}
\caption{Fig.~\newFigure{example1}. Two histograms (blue squares and red circles) to be compared for consistency.}
}

\medskip

There are at least two variants of interest to this question:

\item{1.} We wish to test the hypothesis:
\itemitem{} $H_0$: The means of the two histograms are bin-by-bin equal, against 
\itemitem{} $H_1$: The means of the two histograms are not bin-by-bin equal.  

\item{2.} We wish to test the hypothesis:
\itemitem{} $H_0^\prime$: The densities of the two histograms are bin-by-bin equal, against 
\itemitem{} $H_1^\prime$: The densities of the two histograms are not bin-by-bin equal.

\noindent In the second case, the relative normalization of the two histograms
is not an issue: we only compare the shapes. 

It may be noted that there are a large 
variety of tests that attempt to answer the question of whether a given
dataset is consistent with having been drawn from some specified continuous
distribution. These tests may typically be adapted to address the question
of whether two datasets have been drawn from the same continuous distribution,
often referred to as ``two-sample'' tests. These tests may further be
adapted to the present problem, that of determining whether two histograms
have the same shape. This situation is also referred to as comparing whether
two (or more) rows of a ``table'' are consistent. The datasets of this form are
also referred to as ``grouped data''.

Although we keep the discussion focussed on the comparison of two histograms,
it is worth remarking that many of the observations apply also to other 
situations, such as the comparison of a histogram with a model prediction.

\bigskip

\chapter{ch:notation}{Notation}

\smallskip

We assume that we have formed our two histograms with the same number
of bins, $k$, with identical bin boundaries. The bin contents of the 
``first'' histogram are given by realization $u$ of random variable
$U$, and of the second by realization $v$ of random variable $V$. Thus,
the sampling distributions are:
$$\eqalign{
P(U=u) &= \prod_{i=1}^k {\mu_i^{u_i}\over u_i!}e^{-\mu_i},\cr
P(V=v) &= \prod_{i=1}^k {\nu_i^{v_i}\over v_i!}e^{-\nu_i},\cr
}\eqno{(\newEq{e:nota})}$$
where the vectors $\mu$ and $\nu$ are the mean bin contents of the 
respective histograms.

We define:
$$\eqalign{
  N_u &\equiv \sum_{i=1}^k U_i,\quad\hbox{total contents of first histogram,}\cr
  N_v &\equiv \sum_{i=1}^k V_i,\quad\hbox{total contents of second histogram,}\cr
  \mu_T &\equiv \vev{N_u} = \sum_{i=1}^k \mu_i,\cr
  \nu_T &\equiv \vev{N_v} = \sum_{i=1}^k \nu_i,\cr
  t_i &\equiv u_i + v_i,\quad i=1,\ldots,k.\cr
}$$

We are interested in the power of a test, at any given
confidence level. The power is the probability that the null hypothesis
is rejected when it is false. Of course, the power depends on the
true sampling distribution. In other words, the power is one minus
the probability of a Type II error. The confidence level is the probability
that the null hypothesis is accepted, if the null hypothesis is correct. Thus,
the confidence level is one minus the probability of a Type I error.
In physics, we usually don't specify the confidence level of a test in
advance, at least not formally. Instead, we quote the $P$-value for our
result. This is the probability, under the null hypothesis, of obtaining
a result as ``bad'' or worse than our observed value. This would be the
probability of a Type I error if our observation were used to define the
critical region of the test.

Note that we are dealing with discrete distributions here, and 
exact statements of frequency are problematic, though not 
impossible. Instead of attempting to construct exact statements, our
treatment of the discreteness will be such as to err on the 
``conservative'' side. By ``conservative'', we mean that we will
tend to accept the null hypothesis with greater than the stated
probabitlity. It is important to understand that this is not always
the ``conservative'' direction, for example it could mislead us into
accepting a model when it should be rejected.

We will drop the distinction between the random variable (upper case
symbols $U$ and $V$) and a realization (lower case $u$ and $v$) in the
following, but will point out where this informality may yield confusion.

The computations in this note are carried out in the framework of the
R statistics package [\newCite{R}{R Development Core Team, R: A Language and Environment for Statistical Computing, R Foundation for Statistical Computing, Vienna 2007, ISBN 3-900051-07-0, http://www.r-project.org/.}].

\bigskip

\section{s:LSC}{Large Statistics Case}

\smallskip

If all of the bin contents of both histograms are large, then 
we may use the approximation that the bin contents are normally
distributed.

Under $H_0$,
$$\vev{u_i} = \vev{v_i} \equiv \mu_i,\ i=1,\ldots,k.$$
More properly, it is $\vev{U_i}=\mu_i$, etc., but we are
permitting $u_i$ to stand for the random variable as well as its
realization, as noted above.
Let the difference in the contents of bin $i$ between the two histograms be:
$$\Delta_i \equiv u_i-v_i,$$
and let the standard deviation for $\Delta_i$ be denoted $\sigma_i$.
Then the sampling distribution of the difference between the two histograms
is:
$$P(\Delta) = {1\over (2\pi)^{k/2}}\left(\prod_{i=1}^k {1\over\sigma_i}\right)\exp\left(-{1\over 2}\sum_{i=1}^k {\Delta_i^2\over\sigma_i^2}\right).
$$

This suggests the test statistic:
$$T = \sum_{i=1}^k {\Delta_i^2\over\sigma_i^2}.$$
If the $\sigma_i$ were known, this would simply be 
distributed according to the chi-square distribution with
$k$ degrees of freedom. The maximum-likelihood estimator for
the mean of a Poisson is just the sampled number. The mean of the
Poisson is also its variance, and we will use the sampled number
also as the estimate of the variance in the normal approximation.

We suggest the following algorithm for this test:
\item{1.} For $\sigma_i^2$ form the estimate
$$\hat\sigma_i^2 = (u_i+v_i).$$
\item{2.} Statistic $T$ is thus evaluated according to:
$$T = \sum_{i=1}^k {(u_i-v_i)^2\over u_i+v_i}.$$
If $u_i=v_i=0$ for bin $i$, the contribution to the sum from that
bin is zero.
\item{3.} Estimate the $P$-value according to a chi-square with $k$
degrees of freedom. Note that this is not an exact result.

If it is desired to only compare shapes, then the suggested algorithm is 
to scale both histogram bin contents:
\item{1.} Let
$$N=0.5(N_u+N_v).$$
Scale $u$ and $v$ according to:
$$\eqalign{
  u_i &\to u_i^\prime = u_i(N/N_u)\cr
  v_i &\to v_i^\prime = v_i(N/N_v).\cr
}$$
\item{2.} Estimate $\sigma_i^2$ with:
$$
  \hat\sigma_i^2 = \left({N\over N_u}\right)^2u_i + \left({N\over N_v}\right)^2v_i.
$$
\item{3.}  Statistic $T$ is thus evaluated according to:
$$T = \sum_{i=1}^k {\left({u_i\over N_u}-{v_i\over N_v}\right)^2\over {u_i\over N_u^2}+{v_i\over N_v^2}}.$$
\item{3.} Estimate the $P$-value according to a chi-square with $k-1$
degrees of freedom. Note that this is not an exact result.

Due to the presence of bins with small bin counts, we might not expect this method to be especially good for the data in Fig.~1,
but we can try it anyway. Table~\refno{t:largestat}\
gives the results of applying this test, both including the normalization
and only comparing shapes.

\bigskip

\vbox{
\caption{Table~\newTable{t:largestat}. Results of tests for consistency of the two
datasets in Fig.~\refno{example1}. The tests below the $\chi^2$ lines are 
described in Section 3.}
\hbox{\centerline{\vbox{ 
  \hrule
  \halign{#\ \hfil&#\quad&#\quad&#\quad&#\cr
  \noalign{\vskip3pt}
  Type of test & \hfil$T$\hfil & \hfil NDOF\hfil & \hfil $P(\chi^2>T)$\hfil & \hfil $P$-value\hfil \cr
  \noalign{\vskip3pt\hrule\vskip3pt}
  $\chi^2$ Absolute comparison & 29.8 & 40 & 0.88 & 0.86 \cr
  $\chi^2$ Shape comparison & 24.9 & 39 & 0.96 & 0.95 \cr
  \noalign{\vskip3pt\hrule\vskip3pt}
  Likelihood Ratio Shape comparison & 25.3 & 39 & 0.96 & 0.96\cr
  Kolmogorov-Smirnov Shape comparison & 0.043 & 39 & NA & 0.61\cr
  Bhattacharyya Shape comparison & 0.986 & 39 & NA & 0.97\cr
  Cram\'er-Von-Mises Shape comparison & 0.132 & 39 & NA & 0.45\cr
  Anderson-Darling Shape comparison & 0.849 & 39 & NA & 0.45\cr
  Likelihood value shape comparison & 79 & 39 & NA & 0.91\cr
  }\vskip2pt\hrule
}}}}

\bigskip
  
In the column labeled ``$P$-value'' an attempt is made to compute (by simulation)
a more reliable estimate of the probability, under the null hypothesis, that
a value for $T$ will be as large as that observed. This may be compared with the 
$P(\chi^2>T)$ column, which is the probability assuming $T$ follows a 
$\chi^2$ distribution with NDOF degress of freedom.
  
Note that the absolute comparison yields slightly poorer agreement between the histograms than
the shape comparison. The total number of
counts in one dataset is 492; in the other it is 424. Treating these as samplings
from a normal distribution with variances 492 and 424, we find a 
difference of 2.2 standard deviations or a two-tailed $P$-value of
$0.025$. This low probability is diluted by the bin-by-bin
test. Using a bin-by-bin test to check whether the totals are consistent is not 
a powerful approach. In fact, the two histograms were generated with a 10\% difference 
in expected counts.
  
The evaluation by simulation of the probability under the null hypothesis is in fact
problematic, since the null hypothesis actually isn't completely specified.
The problem is the dependence of Poisson probabilities on the absolute
numbers of counts. Probabilities for differences in Poisson counts are
not invariant under the total number of counts. Unfortunately, we don't 
know the true mean numbers of counts in each bin. Thus, we must estimate
these means. The procedure adopted here is to use the maximum likelihood estimators (see below) for the mean numbers, in the null hypothesis. We'll have further
discussion of this procedure below -- it does not always yield valid results.
  
In our example, the probabilities estimated according to our
simulation and the probabilities according to a $\chi^2$ distribution are close to each other. This suggests the possibility of using the
$\chi^2$ probabilities -- if we can do this, the problem that we haven't
completely specified the null hypothesis is avoided. We offer the following
conjecture:

\noindent {\bf Conjecture:} Let $T$ be the test statistic described above, for either
the absolute or the shape comparison, as desired. Let $T_c$ be a possible
value of $T$ (perhaps the critical value to be used in a hypothesis test).
Then, for large values of $T_c$:
$$
  P(T<T_c) \ge P\left(T<T_c | \chi^2(T,\hbox{ndof})\right),
$$
where $P\left(T<T_c | \chi^2(T,\hbox{ndof})\right)$ is the probability that $T<T_c$
according to a $\chi^2$ distribution with ndof degrees of freedom (either 
$k$ or $k-1$, according to which test is being performed).  
  
We'll only suggest an approach to a proof, which could presumably also be used
to develop a formal
condition for $T_c$ to be ``large''. The conjecture also appears to
be true anecdotally, and for interesting values of $T_c$, noting that it
is large values of $T_c$ that we care most about for an interesting 
hypothesis test. 

We provide some intuition for the conjecture
by considering the case of one bin. For simplicity we'll also suppose
that $\nu=\mu$ and that $\mu$ is small ($\ll 1$, say). Since we are 
interested in large values of the statistic, we are interested in 
the situation where one of $u,v$ is large, and the other small (since
$\mu$ is small).
Suppose it is $u$ that is large. Then
$$
 T={(u-v)^2\over u+v} \approx u.
$$
For given $v$ (0, say), the probability of $T$ is thus
$$
  P(T)\approx {\mu^T\over T!}e^{-\mu}.
$$
This may be compared with the chi-square probability distribution
for one degree of freedom:
$$
  P(T=\chi^2) = {1\over \sqrt{2\pi}}{e^{-T/2}\over\sqrt{T}}.
$$
The ratio is, dropping the constants:
$$
{P(T)\over P(T=\chi^2)} \propto {\mu^T e^{T/2}\sqrt{T}\over T!} = {\exp\left[T\left({1\over 2}+\ln\mu\right)\right]\over \sqrt{T}\Gamma(T)},
$$
which approaches zero for large $T$, for any given $\mu$.
We conclude that the conjecture is valid in the case of one bin, and strongly suspect that
the argument generalizes to multiple bins.
 
According to the conjecture, if we use the
probabilities from a $\chi^2$ distribution in our test, the error that we
make is in the ``conservative'' direction (as long as $T_c$ is large). That is, we'll reject the
null hypothesis less often than we would with the correct probability.
It should be emphasized that this conjecture is independent of the
statistics of the sample, bins with zero counts are fine. In the
limit of large statistics, the inequality approaches equality.

Lest we conclude that it is acceptable to just use this great
simplification in all situations, we hasten to point out that it isn't
as nice as it sounds. The problem is that, in low statistics situations,
the power of the test according to this approach can be dismal. That is,
we might not reject the null hypothesis in situations where it is 
obviously implausible.

We may illustrate these considerations with some simple examples, see Fig.~\refno{chisqConj}. The plot for high
statistics on the left shows excellent agreement between the actual distribution and the $\chi^2$ distribution.
The lower statistics plots in the middle and right, for two different models, show that the chi-square
approximation is very conservative in general. Thus, using the chi-square probability lacks power in this case, and
is not a recommended approximation.

\vbox{
\hbox{\centerline{\includegraphics[width=6in]{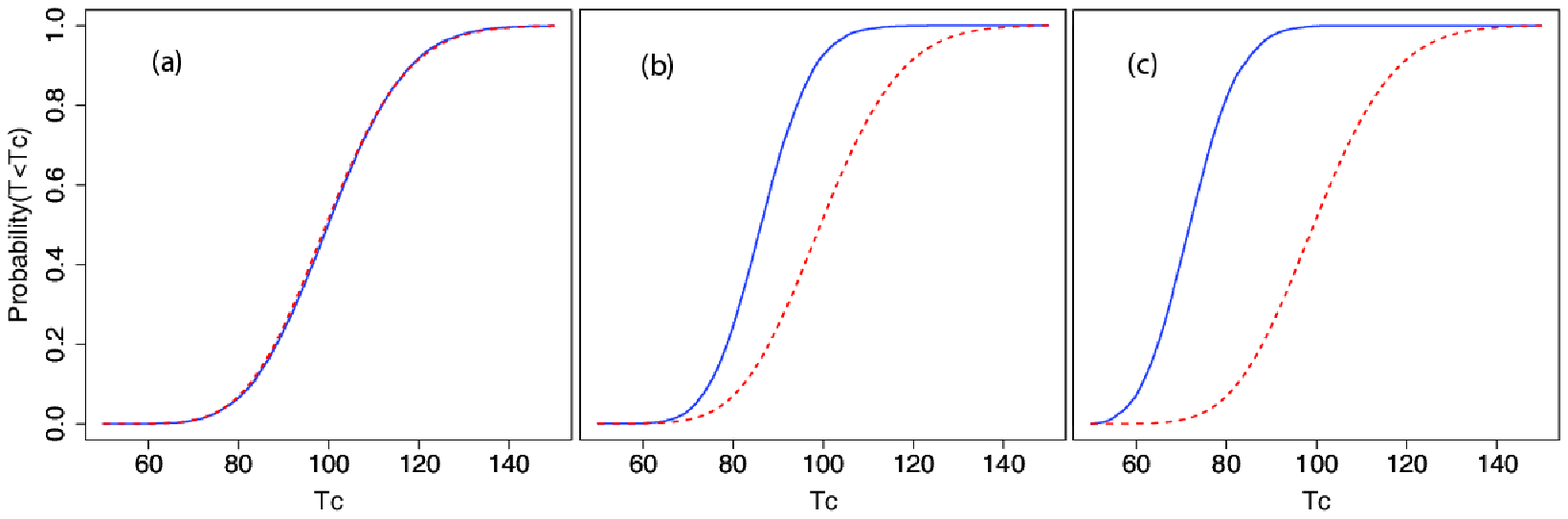}}}
\vskip-.1cm
\caption{Fig.~\newFigure{chisqConj}. Comparison of the actual (cumulative) probability distribution for $T$ with the
chi-square distribution. The solid blue curves show the actual distributions, and the dashed
red curves the chi-square distributions. All plots are for 100 bin histograms.
(a) Each bin has mean 100. (b) Each bin has mean 1. (c) Bin $j$ has mean $30/j$.} 
}
  
\bigskip

\chapter{ch:generalCase}{General Case}

\smallskip

If the bin contents are not necessarily large, then the normal
approximation may not be good enough. There are various approaches we
could take in this case. We'll discuss and compare several possibilities.

\vfil\break

\section{s:CB}{Combining Bins}

A simple approach is to combine bins until the normal 
approximation is good enough. In many cases this doesn't lose too
much statistical power. It may be necessary to check with simulations that 
probability statements are valid. Figure~\refno{f:emuChisqTk} shows the results of this approach on the data in Figure~\refno{example1}, as a function of the minimum number of events per bin. The comparison being made is for the shapes.
The algorithm is to combine corresponding bins in both histograms until both have at least ``{\tt minBin}'' counts in each bin.

\medskip

\vbox{\hbox{%
\centerline{\includegraphics[width=6.1in]{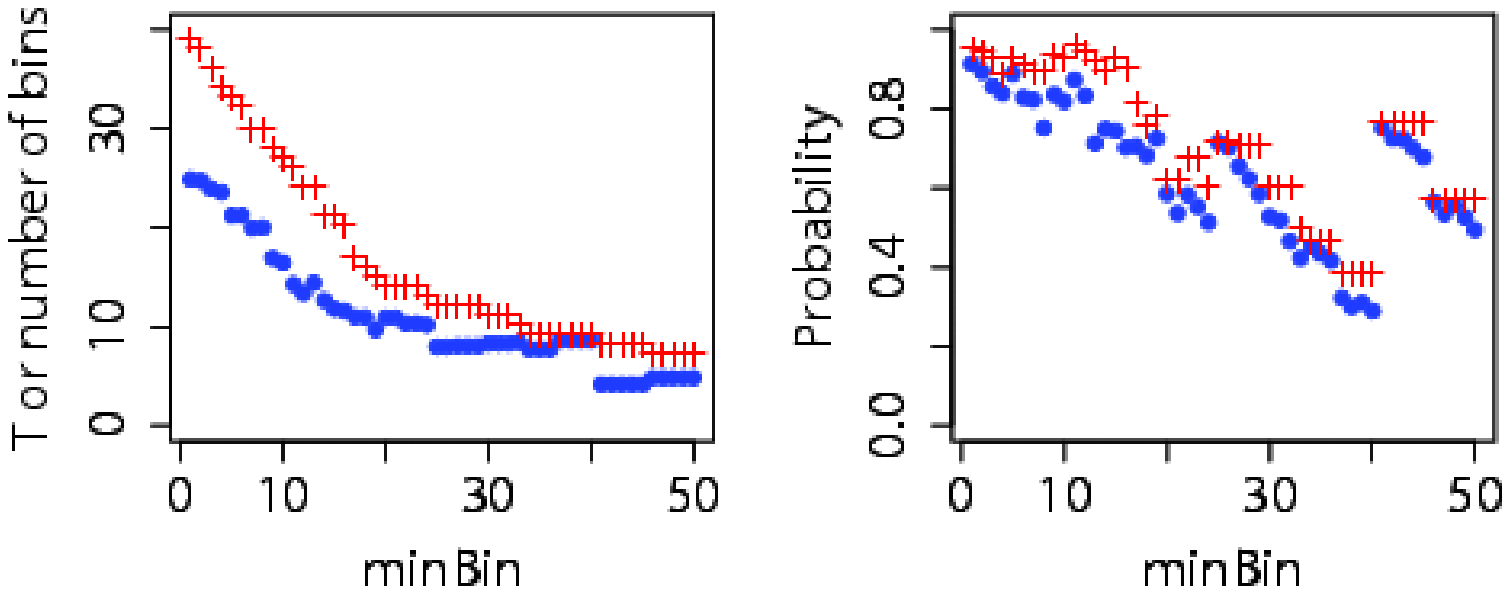}}}
\vskip-12pt
\caption{Fig.~\newFigure{f:emuChisqTk}. Left: The blue dots show the value of the test statistic $T$, and the red pluses shows the number of histogram bins for the data in Fig.~1,
as a function of the minimum number of counts per histogram bin. 
Right: The $P$-value for consistency of the two datasets in Fig.~\refno{example1}.
The red pluses show the probability for a chi-square distribution, and the blue circles
show the probability for the actual distribution, with an estimated
null hypothesis.}
}

\section{s:TEN}{Testing for Equal Normalization}

An alternative is to work with the Poisson distributions. Let us separate the
problem of the shape from the problem of the overall normalization. In the case
of testing equality of overall normization,
there is a well-motivated choice for the test statistic, even for low statistics. 

To test the
normalization, we simply compare totals over all bins between the two 
histograms. Our distribution is
$$
  P(N_u,N_v) = {\mu_T^{N_u}\nu_T^{N_v}\over N_u!N_v!}e^{-(\mu_T+\nu_T)}.
$$
The null hypothesis is $H_0:\mu_T = \nu_T$, to be tested against
alternative  $H_1:\mu_T \ne \nu_T$. We are thus interested in the 
difference between the two means; the sum is effectively a nuisance
parameter. That is, we are interested in
$$\eqalignno{
 P(N_v | N_u+N_v = N) &= {P(N | N_v) P(N_v) \over P(N)}\cr
   &= {\mu_T^{N-N_v}e^{-\mu_T}\over (N-N_v)!}{\nu_T^{N_v}e^{-\nu_T}\over N_v!}\Bigg/
      {(\mu_T+\nu_T)^N e^{-(\mu_T+\nu_T)}\over N!}\cr
   &= \pmatrix{N\cr N_v} \left({\nu_T \over \mu_T + \nu_T}\right)^{N_v} 
         \left({\mu_T \over \mu_T + \nu_T}\right)^{N-N_v}.\cr
}$$            
This probability now permits us to construct a uniformly most
powerful test of our hypothesis (Ref.~\newCite{b:Lehmann}{E.~L.~Lehmann and Joseph P.~Romano, {\sl Testing Statistical Hypotheses}, Third edition, Springer, New York (2005), Theorem 4.4.1.}). Note that it is simply a binomial
distribution, for given $N$. The uniformly most powerful property holds
independently of $N$, although the probabilities cannot be computed without
$N$.

The null hypothesis corresponds to $\mu_T=\nu_T$, that is:
$$P(N_v | N_u+N_v = N) = \pmatrix{N\cr N_v} \left({1\over 2}\right)^N.
$$         
For our example, with $N=916$ and $N_v=424$, the $P$-value is
$0.027$, assuming a two-tailed probability is desired.
This may be compared with our earlier estimate of $0.025$ in the
normal approximation. Note that for our binomial calculation we have 
``conservatively'' included the endpoints (424 and 492). If we try to mimic
more closely the normal estimate by subtracting one-half the probability
at the endpoints, we obtain $0.025$, essentially the
normal number we found earlier. The {\tt dbinom} function~Ref.~\newCite{b:Rdbinom}{%
http://www.herine.net/stat/software/dbinom.html.} in the R package has
been used for this computation.

\vfil\break

\section{s:SCS}{Shape Comparison Statistics}

There are many different possible statistics for comparing the 
shapes of the histograms. We investigate several choices. 
Table~\refno{t:largestat} summarizes the result of each of these
tests applied to the example in Fig.~\refno{example1}. We list the
statistics here, and discuss performance in the following sections.


\subsection{s:Chi}{Chi-square test for shape}

Even though we don't expect it to follow a $\chi^2$
distribution, we may evaluate the test statistic:
$$\chi^2 = \sum_{i=1}^k {\left({u_i\over N_u}-{v_i\over N_v}\right)^2\over {u_i\over N_u^2}+{v_i\over N_v^2}}.$$
If $u_i=v_i=0$, the contribution to the sum from that bin is zero. We have already discussed application of
this statistic to the example of Fig.~\refno{example1}.

\subsection{s:BDM}{Geometric test for shape}

Another test statistic we could try may be motivated from a geometric
perspective. We consider the bin contents of a histogram to define a
vector in a $k$-dimensional space. If two such vectors are drawn from the same
distribution (the null hypothesis), then they will tend to point in the same
direction (we are not interested in the lengths of the vectors here).
Thus, if we represent each histogram as a unit vector with components:
$$\{u_1/N_u,\ldots,u_k/N_u\}, \hbox{ and } \{v_1/N_v,\ldots,v_k/N_v\},$$
we may form the test statistic:
$$T_{\rm BDM}= \sqrt{{u\over N_u}\cdot {v\over N_v}} = \left(\sum_{i=1}^k {u_iv_i\over N_u N_v}\right)^{1/2}.$$
This is known as the ``Bhattacharyya distance measure''. We'll refer to it
as the ``BDM'' statistic for short. 
We assume that neither histogram is empty for this statistic.
All vectors lie in the positive direction in all coordniates, so
there is no issue with taking the square root.

It may be noticed that this statistic is related to the $\chi^2$ statistic -- the
${u\over N_u}\cdot {v\over N_v}$ dot product is close to the cross term in the $\chi^2$
expression.

We apply this formalism to the example in Fig.~\refno{example1}. The resulting
terms in the sum over bins are shown in Fig.~\refno{f:sqrtemu}. The sum over bins 
gives 0.986 (See Table~\refno{t:largestat}\ for a summary). According to our estimated distribution
of this statistic under the null hypothesis, this gives a $P$-value of 0.97, 
similar to the $\chi^2$ test result. 

\noindent\includegraphics[width=2.9in]{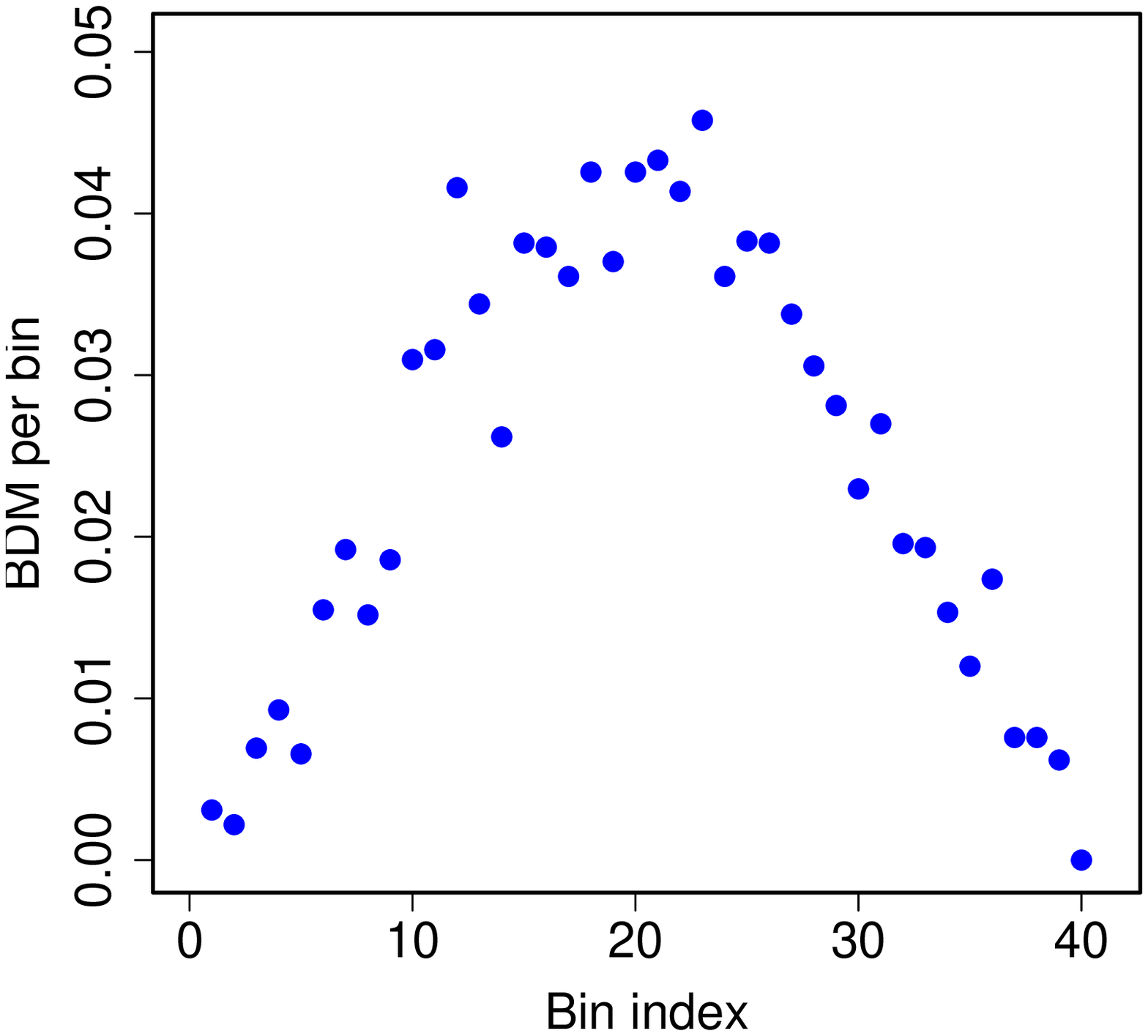}
\includegraphics[width=3in]{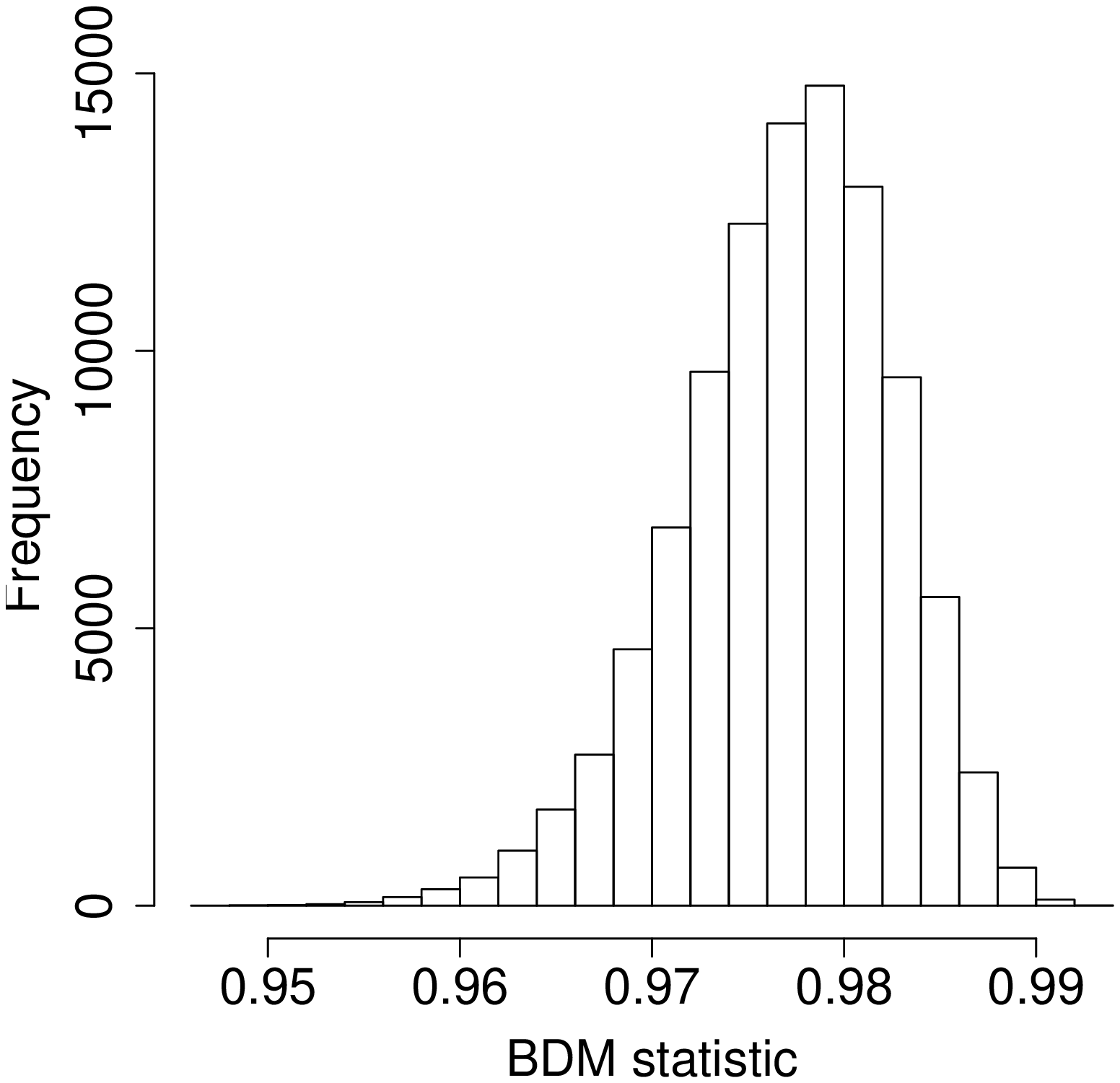}
\caption{Fig.~\newFigure{f:sqrtemu}. Left: Bin-by-bin contributions to the geometric (``BDM'') test statistic 
for the example of Fig.~\refno{example1}.
Right: Estimated distribution of the BDM statistic
for the null hypothesis in the example of Fig.~\refno{example1}.}

\subsection{s:KS}{Kolmogorov-Smirnov test}

Another approach to a shape test may be based on the Kolmogorov-Smirnov (KS)
idea. Recall that the idea of the KS test is to estimate the maximum 
difference between observed and predicted cumulative distribution functions (CDFs) and compare with 
expectations. We may adapt this idea to the present case. 
It should be remarked that if we have the actual data points from which the histograms
are derived, then we may use the Kolmogorov-Smirnov (``KS'') procedure
directly on those points. This would incorporate additional information
and yield a potentially more powerful test. However, if the bin widths are
small compared with possible structure it may be expected to not make much 
difference.

We modify the KS statistic to apply to comparison of histograms as follows.
We assume that neither histogram is empty. Form the ``cumulative distribution histograms''
according to:

\vskip-10pt

\noindent$$\eqalign{u_{ci} &= \sum_{j=1}^i u_j/N_u\cr
           v_{ci} &= \sum_{j=1}^i v_j/N_v.\cr
}$$

\vskip-10pt

\noindent Then compute the test statistic:

\vskip-10pt

\noindent $$T_{\rm KS} = \max_i |u_{ci}-v_{ci}|.$$
Test statistics may also be formed for one-tail tests, but
we consider only the two-tail test here.           

We apply this formalism to the example in Fig.~\refno{example1}. The bin-by-bin distances
are shown in Fig.~\refno{f:KSemu}. The maximum over bins 
gives 0.043 (See Table~\refno{t:largestat}\ for a summary). According to our estimated distribution
of this statistic under the null hypothesis, this gives a $P$-value of 0.61, somewhat 
smaller than for the $\chi^2$ test result, but still indicating consistency
of the two histograms. Note that the KS test tends to emphasize the
region near the peak of the distribution, that is the 
region where the largest fluctuations are expected in Poisson statistics. 

\vskip-12pt

\vbox{%
\noindent\includegraphics[width=5.9in]{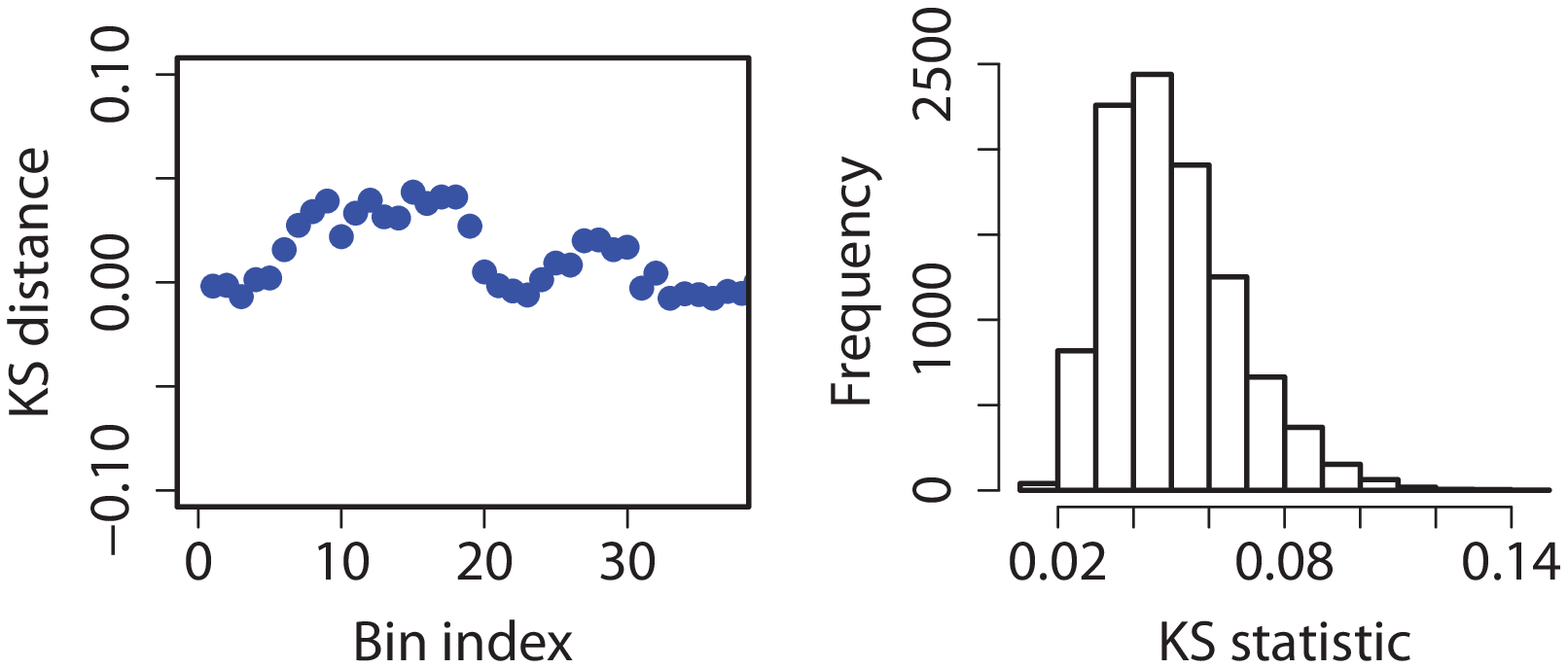}
\caption{Fig.~\newFigure{f:KSemu}. Left: Bin-by-bin distances for the Kolmogorov-Smirnov test statistic 
for the example of Fig.~\refno{example1}.
Right: Estimated distribution of the Kolmogorov-Smirnov distance
for the null hypothesis in the example of Fig.~\refno{example1}.} 
}

\subsection{s:CVM}{Cram\'er-von-Mises test}

Somewhat similar to the Kolmogorov-Smirnov test is the Cram\'er-von-Mises (CVM)
test. The idea in this test is to add up the squared differences between the cumulative
distributions being compared. Again, this test is usually thought of as a test to compare
an observed distribution with a presumed parent continuous 
probability distribution. However, the algorithm can be 
adapted to the two-sample comparison, and to the case of comparing
two histograms. 

The test statistic for comparing the two samples $x_1,x_2,\ldots,x_N$ and
$y_1,y_2,\ldots,y_M$ is [\newCite{b:Anderson62}{T.~W.~Anderson, {\sl On the Distribution of the
Two-Sample Cram\'er-Von Mises Criterion}, Annals Math.\ Stat., {\bf 33} (1962) 1148.}]:
$$T = {NM\over (N+M)^2}\left\{\sum_{i=1}^N\left[E_x(x_i)-E_y(x_i)\right]^2
  + \sum_{j=1}^M\left[E_x(y_j)-E_y(y_j)\right]^2\right\},$$
where $E_x$ is the empirical cumulative distribution for sampling $x$. That is,
$E_x(x) = n/N$ if $n$ of the sampled $x_i$ are less than or equal to $x$.

We adapt this for the present application of comparing histograms with bin contents
$u_1, u_2,\ldots,u_k$ and $v_1, v_2,\ldots,v_k$ with identical bin boundaries:
Let $z$ be a point in bin $i$, and define the empirical cumulative distribution function
for histogram $u$ as:
$$E_u(z) = \sum_{j=1}^i u_i/N_u.$$
Then the test statistic is:
$$
T_{\rm CVM} = {N_uN_v\over (N_u+N_v)^2} \sum_{j=1}^k (u_j+v_j)\left[E_u(z_j)-E_v(z_j)\right]^2.
$$

We apply this formalism to the example in Fig.~\refno{example1}, finding $T_{\rm CVM} = 0.132$. The resulting
estimated distribution under the null hypothesis is shown in Fig.~\refno{f:CVMADemu}.   According to our estimated distribution
of this statistic under the null hypothesis, this gives a $P$-value of 0.45 (See Table~\refno{t:largestat}\ for a summary), somewhat  
smaller than the $\chi^2$ test result.

\noindent\includegraphics[width=2.9in]{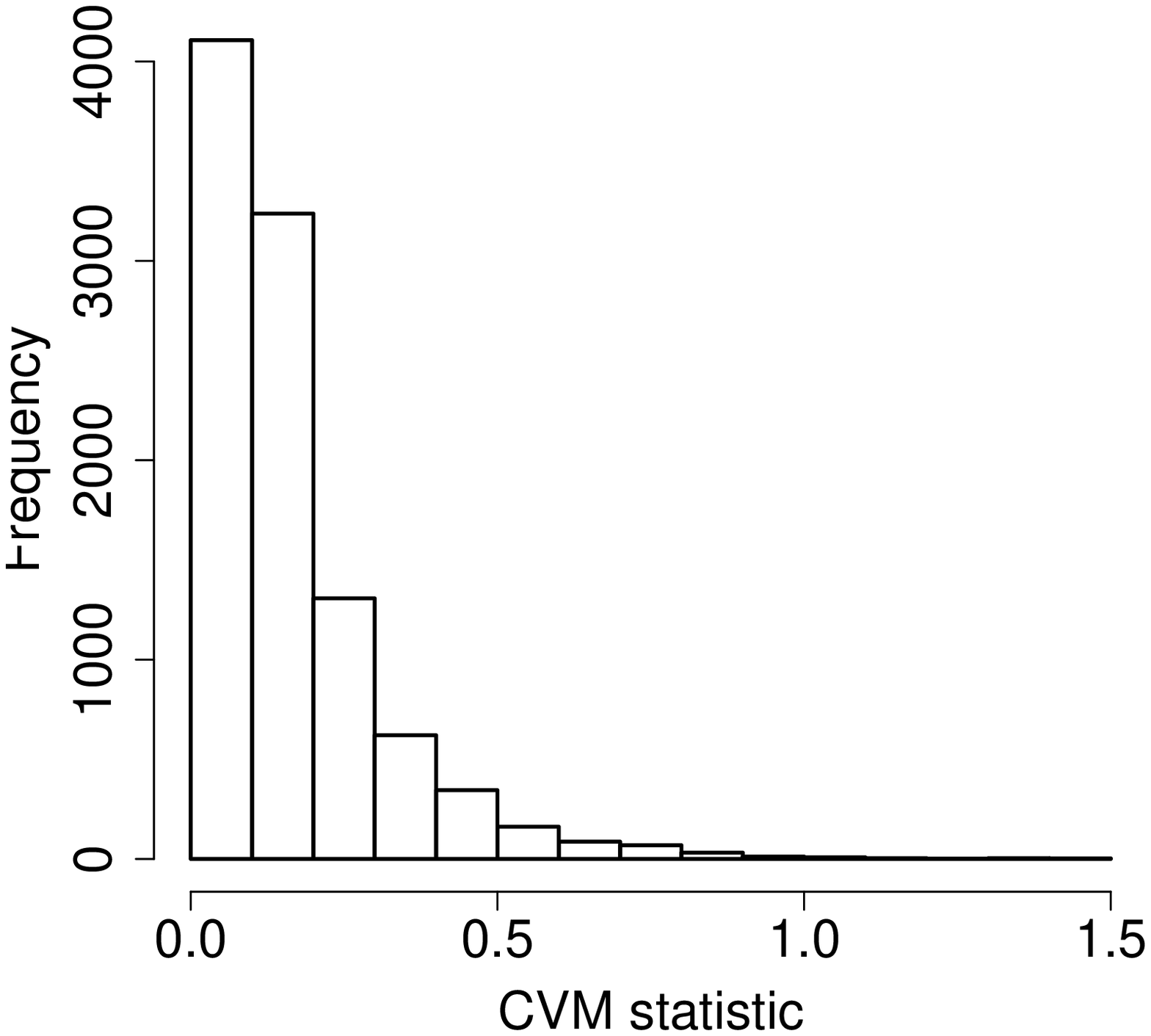}
\includegraphics[width=3in]{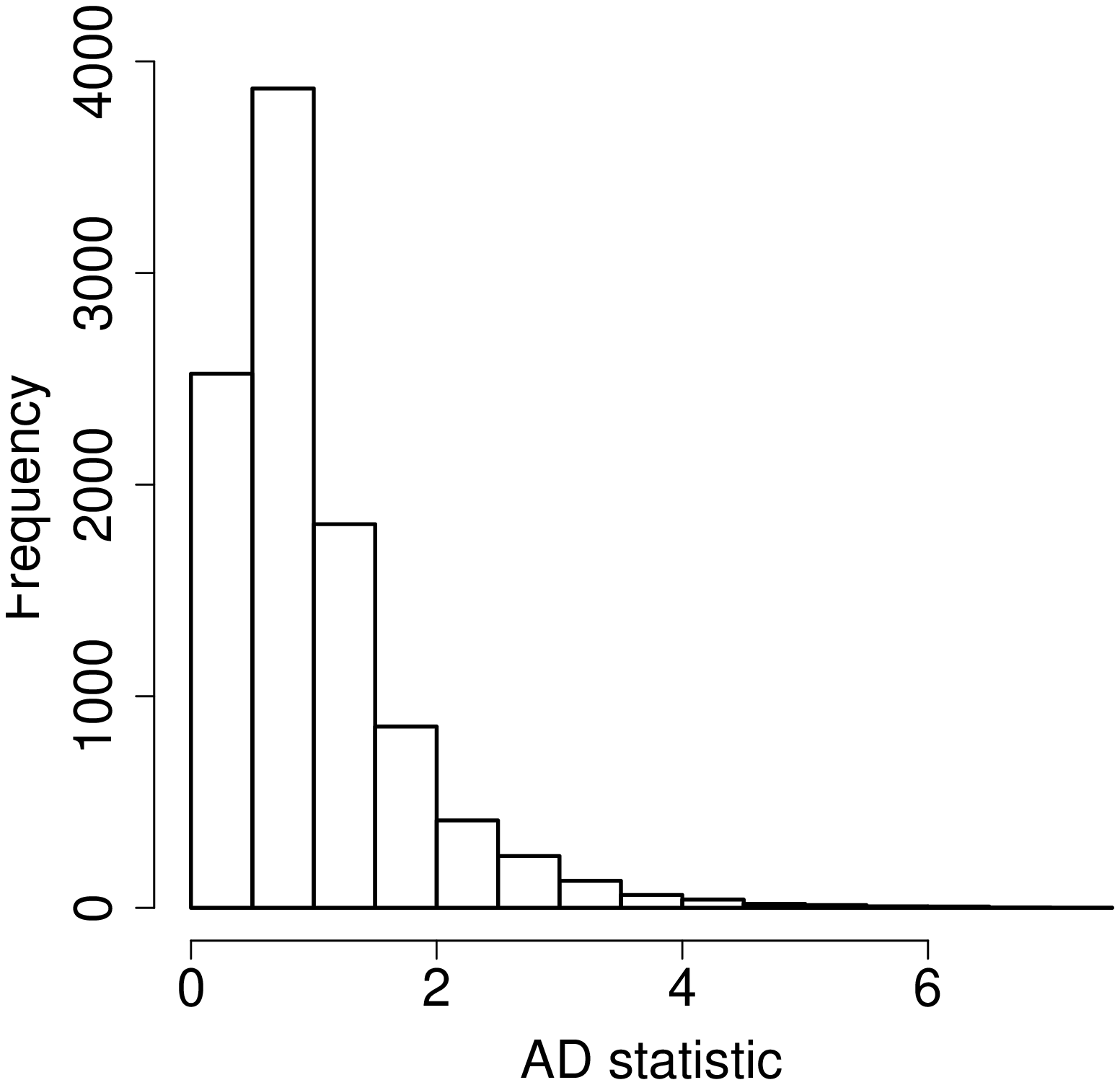}
\caption{Fig.~\newFigure{f:CVMADemu}. Estimated distributions of the test 
statistic for the null hypothesis in the example of Fig.~\refno{example1}
Left: The Cram\'er-von-Mises statistic.
Right: The Anderson-Darling statistic.} 

\subsection{s:AD}{Anderson-Darling test for shape}

The Anderson-Darling test is another variant on the theme of non-parametric comparison of cumulative distributions. It is similar to the Cram\'er-von-Mises statistic, but is designed to
be sensitive to the tails of the CDF. The original statistic was, once again, designed to compare
a dataset drawn from a continuous distribution, with CDF $F_0(x)$ under the null hypothesis:
$$
 A_m^2 = m \int_{-\infty}^\infty {\left[F_m(x) - F_0(x)\right]^2\over F_0(x)\left[1-F_0(x)\right]}dF_0(x),
$$
where $F_m(x)$ is the empirical CDF of dataset $x_1,\ldots x_m$.   
Scholz and Stephens~[\newCite{b:scholz}{F.~W.~Scholz and M.~A.~Stephens, {\sl $k$-Sample Anderson-Darling Tests}, J.~Amer.\ Stat.\ Assoc. {\bf 82} (1987) 918.}] provide a form
of this statistic for a $k$-sample test on grouped data (e.g., as might be used to
compare $k$ histograms). Based on the result expressed in their Eq.~6, the expression of interest to us for two
histograms is:
$$
 \eqalign{
 T_{\rm AD} = {1\over N_u+N_v}\sum_{j=k_{\rm min}}^{k_{\rm max}-1} {t_j\over \Sigma_j\left(N_u+N_v-\Sigma_j\right)}
   \Bigg\{&\left[(N_u+N_v)\Sigma_{uj} - N_u\Sigma_j\right]^2/N_u\cr
  +&\left[(N_u+N_v)\Sigma_{vj} - N_v\Sigma_j\right]^2/N_v\Bigg\},\cr
 }
$$
where $k_{\rm min}$ is the first bin where either histogram has non-zero counts, $k_{\rm max}$ is the number of bins counting up
the the last bin where either histogram has non-zero counts, and
$$\eqalign{
\Sigma_{uj}&\equiv \sum_{i=1}^j u_i,\cr
\Sigma_{vj}&\equiv \sum_{i=1}^j v_i,\quad\hbox{and}\cr
\Sigma_j &\equiv \sum_{i=1}^j t_i = \Sigma_{uj}+\Sigma_{vj}.\cr
}$$
  
We apply this formalism to the example in Fig.~\refno{example1}. The resulting
estimated distribution under the null hypothesis is shown in Fig.~\refno{f:CVMADemu}. The sum over bins 
gives 0.849 (See Table~\refno{t:largestat}\ for a summary). According to our estimated distribution
of this statistic under the null hypothesis, this gives a $P$-value of 0.45, somewhat  
smaller than the $\chi^2$ test result, but similar with the CVM result.

\subsection{s:LR}{Likelihood ratio test for shape}
   
We may base a test whether the histograms are sampled from the same
shape distribution on the same binomial idea as we used for the
normalization test. In this case, however,
there is a binomial associated with each bin of the histogram. We start
with the null hypothesis, that the two histograms are sampled
from the joint distribution:
$$
  P(u,v) = \prod_{i=1}^k {\mu_i^{u_i}\over u_i!}e^{-\mu_i}  
                         {\nu_i^{v_i}\over v_i!}e^{-\nu_i},
$$
where $\nu_i=a\mu_i$ for $i=1,2,\ldots,k$. That is, the ``shapes''
of the two histograms are the same, although the total contents
may differ.     

With $t_i=u_i+v_i$, and fixing the $t_i$ at the observed values
, we have the multi-binomial form:
$$
  P(v|u+v=t) = \prod_{i=1}^k \pmatrix{t_i\cr v_i}\left({\nu_i \over \nu_i + \mu_i}\right)^{v_i}  
       \left({\mu_i \over \nu_i + \mu_i}\right)^{t_i-v_i}.                           
$$
The null hypothesis is that $\nu_i=a\mu_i$ for all values of $i$. We would 
like to test this, but there are now two complications:
\item{1.} The value of ``$a$'' is not specified;
\item{2.} We still have a multivariate distribution. 

For $a$, we will substitute an estimate from the data, namely the maximum
likelihood estimator:
$$\hat a = {N_v\over N_u}.$$
Note that this estimate is a random variable; its use will reduce the
effective number of degrees of freedom by one.

We propose to use a likelihood ratio statistic to reduce the
problem to a single variable. This will be the likelihood under the
null hypothesis (with $a$ given by its maximum likelihood estimator),
divided by the maximum of the likelihood under the alternative 
hypothesis. Thus, we form the ratio:
$$\eqalign{
  \lambda &= {\max_{H_0} {\cal L}(a|v;u+v=t) \over \max_{H_1}{\cal L}(\{a_i\equiv \nu_i/\mu_i\}|v;u+v=t)}\cr
          &= \prod_{i=1}^k {\left({\hat a\over 1+\hat a}\right)^{v_i}\left({1\over 1+\hat a}\right)^{t_i-v_i} \over
          \left({\hat a_i\over 1+\hat a_i}\right)^{v_i}\left({1\over 1+\hat a_i}\right)^{t_i-v_i}}.\cr
}$$          
The maximum likelihood estimator, under $H_1$, for $a_i$ is just
$$
  \hat a_i = v_i/u_i.
$$
Thus, we rewrite our test statistic according to:
$$
  \lambda = \prod_{i=1}^k \left({1+v_i/u_i\over 1+N_v/N_u}\right)^{t_i}
     \left({N_v\over N_u}{u_i\over v_i}\right)^{v_i}.
$$
In practice, we'll work with
$$
 -2\ln\lambda = -2\sum_{i=1}^k \left[t_i \ln \left({1+v_i/u_i\over 1+N_v/N_u}\right) 
  + v_i \ln  \left({N_v\over N_u}{u_i\over v_i}\right) \right].     
$$

Before attempting to apply this, we investigate how
to handle zero bin contents. It is possible that $u_i=v_i=0$ for some bin.
In this case, $P(v_i | u_i+v_i = t_i) =1$, under both $H_0$ and $H_1$, and this bin contributes zero to the sum. It is also possible that $t_i\neq 0$, but
$v_i=0$ or $u_i=0$. If $v_i=0$, then
$$P(0|t_i) = \left({\mu_i\over \nu_i+\mu_i}\right)^{t_i}.$$
Under $H_0$, this is
$$\left({1\over 1+a}\right)^{t_i},$$
and under $H_1$ it is
$$\left({1\over 1+a_i}\right)^{t_i}.$$
The maximum likelihood estimator for $a_i$ is $\hat a_i =0$. Thus,
the likelihood ratio for bin $i$ is
$$
 \lambda_i = \left({1\over 1+\hat a}\right)^{t_i},
$$
and this contributes to the sum an amount:
$$
 -2\ln\lambda_i = -2t_i\ln\left({N_u\over N_u+N_v}\right).$$
If instead $u_i=0$, then   
$$P(t_i|t_i) = \left({\nu_i\over \nu_i+\mu_i}\right)^{t_i}.$$
and the contribution to the sum is
$$
 -2\ln\lambda_i = -2t_i\ln\left({N_v\over N_u+N_v}\right).$$

We apply this formalism to the example in Fig.~\refno{example1}. The resulting
terms in the sum over bins are shown in Fig.~\refno{f:llvscfByBinemu}. The sum over bins 
gives 25.3 (See Table I for a summary). This statistic should 
asymptotically be distributed according to a $\chi^2$ distribution
with the number of degrees of freedom equal to one less than the
number of bins, or $N_{{\rm DOF}} = 39$ in this case. If valid, this
gives a $P$-value of 0.96 in this example. This may be compared with a probability
of 0.96 according to the estimated actual distribution. In this example we obtain 
nearly the same answer as the naive application of the chi-square
calculation with no bins combined. 

\bigskip

\vbox{
\centerline{\includegraphics[width=3.5in,angle=0]{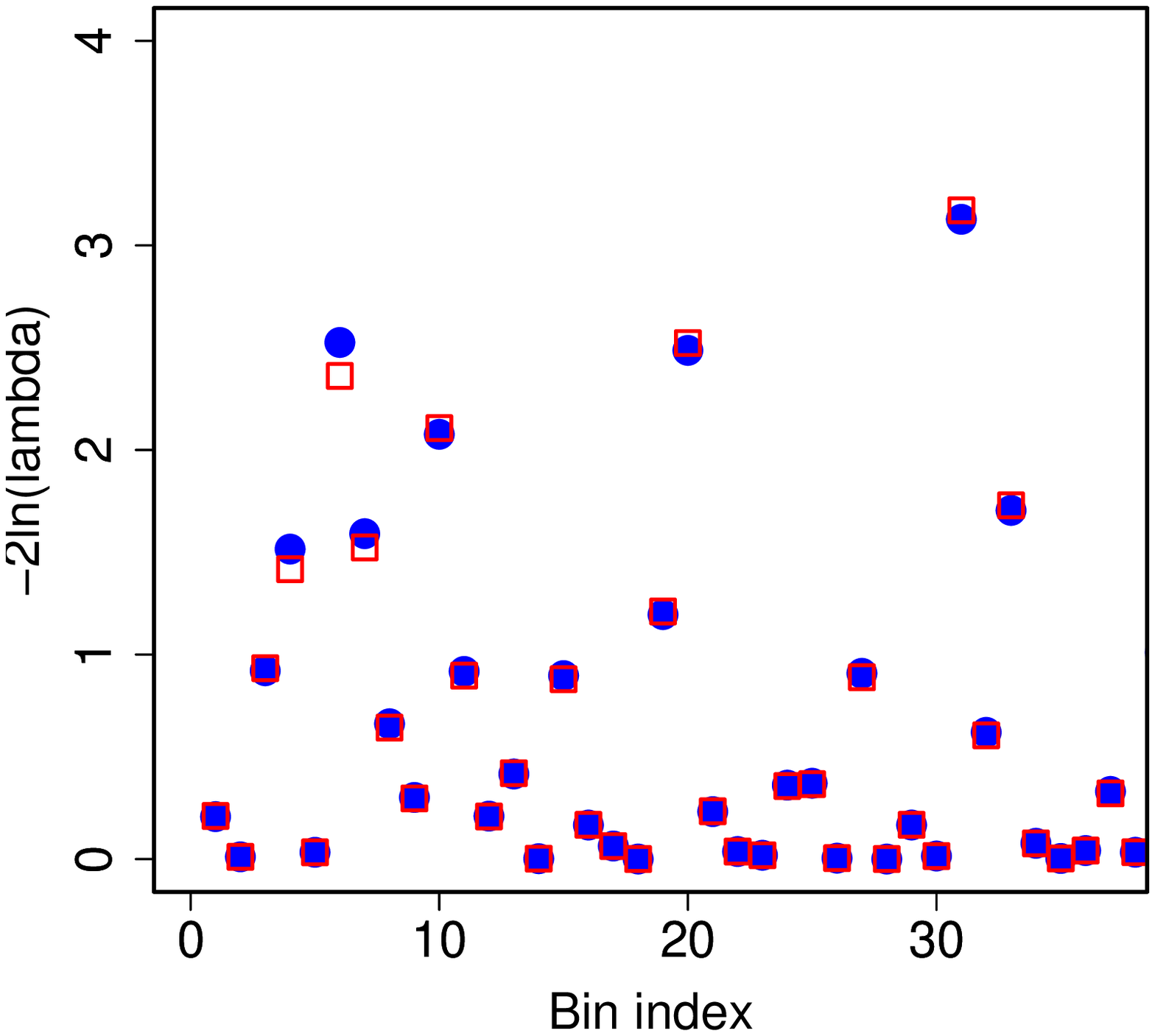}}
\caption{Fig.~\newFigure{f:llvscfByBinemu}. Value of $-2\ln\lambda_i$ or $\chi^2_i$ as a function of histogram bin in the comparison of the two distributions of Fig.~1. Blue circles are $-2\ln\lambda_i$; red squares are $\chi^2_i$.}
}

We may see that this close agreement is a result of nearly bin-by-bin
equality of the two statistics, see Fig.~\refno{f:llvscfByBinemu}. To investigate 
when this might hold more generally, we 
compare the values of $-2\ln\lambda_i$ and $\chi^2_i$ as a function
of $u_i$ and $v_i$, Fig.~\refno{f:llvschisqBin}. 
We observe that the two statistics agree
when $u_i=v_i$ with increasing difference away from that point. This
observation is readily verified analytically. 
This agreement holds even for low statistics. However, we shouldn't 
conclude that the chi-square approximation may be used for low 
statistics -- fluctuations away from equal numbers lead to quite 
different results when we get into the tails at low statistics. 
Our example doesn't really sample these tails.

The precise value of the probability should not be 
taken too seriously, except to conclude that the two distributions are
consistent according to these tests. For example, when we combine bins to 
improve expected $\chi^2$
behavior, we see fairly large fluctuations in the probability estimate just
due to the re-binning (Fig.~\refno{f:emuChisqTk}).

\medskip

\vbox{
\centerline{\includegraphics[width=4.5in]{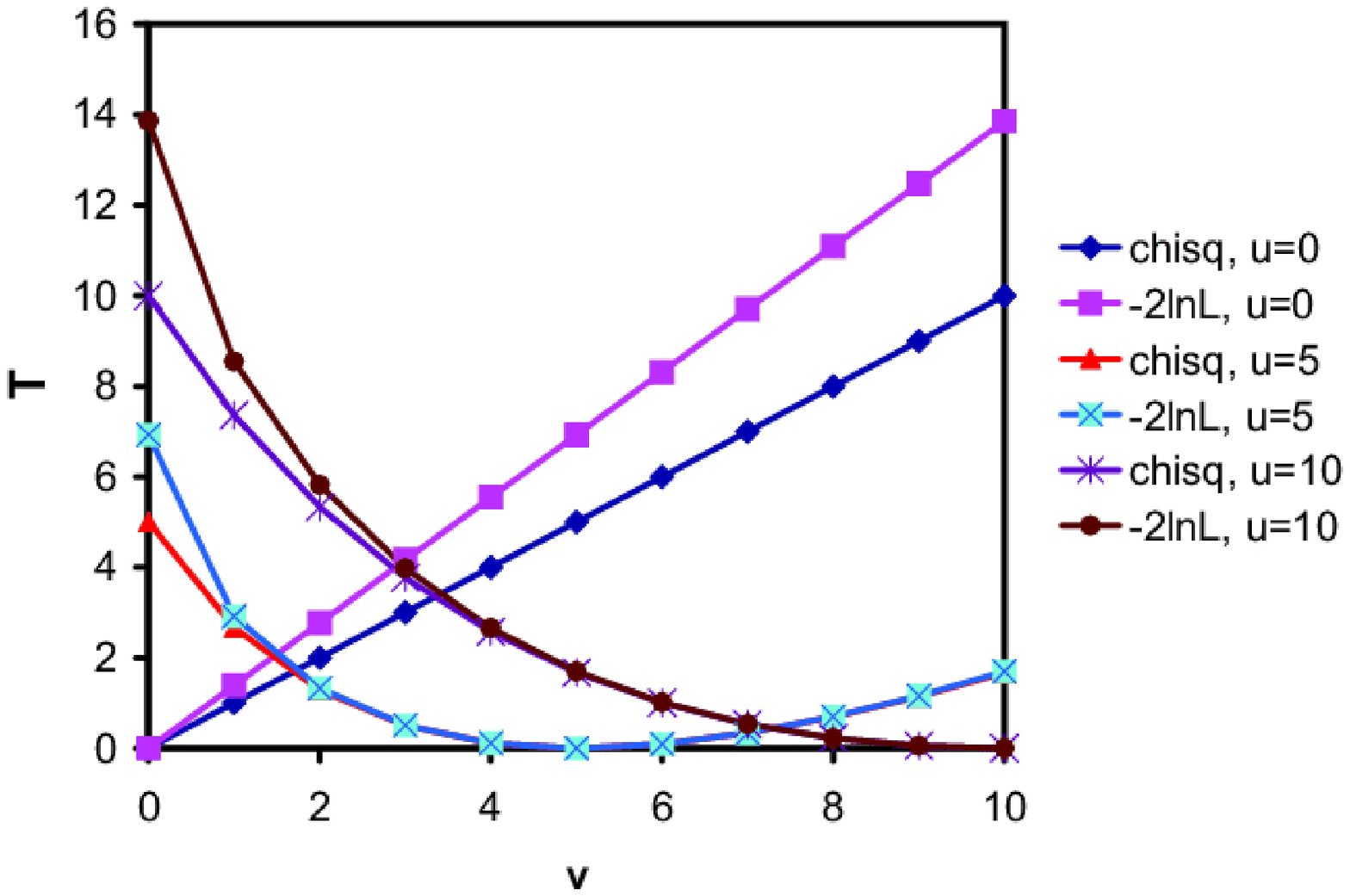}}
\caption{Fig.~\newFigure{f:llvschisqBin}. Value of $-2\ln\lambda_i$ or $\chi^2_i$ as a function of $u_i$ and
$v_i$ bin contents. This plot assumes $N_u=N_v$. The $i$ subscript is
dropped, with the understanding that this comparison is for a single bin.}
}

\subsection{s:LV}{Likelihood value test for shape}
   
An often-used but controversial goodness-of-fit statistic is the
value of the likelihood at its maximum value under the null hypothesis.
It can be demonstrated that this statistic carries little or no information
in some situations. However, in the limit of large statistics it is essentially
the chi-square statistic, so there are known situations were it is a 
plausible statistic to use. We thus look at it here. 

Using the results in the previous section, the test statistic is:
$$
T = -\ln{\cal L} = -\sum_{i=1}^k\left[ \ln\pmatrix{t_i\cr v_i} + t_i\ln {N_u\over N_u+N_v}
    + v_i\ln{N_v\over N_u}\right].
$$
If either $N_u=0$ or $N_v=0$, then $T=0$.
       
We apply this formalism to the example in Fig.~\refno{example1}. The resulting
estimated distribution under the null hypothesis is shown in Fig.~\refno{f:Lemu}. The sum over bins 
gives 90 (See Table~\refno{t:largestat}\ for a summary). According to our estimated distribution
of this statistic under the null hypothesis, this gives a $P$-value of 0.29, similar 
to the $\chi^2$ test result. The fact that it is similar may be expected
from the fact that our example is reasonably well-approximated
by the large statistics limit.

\vbox{
\centerline{\includegraphics[width=3in]{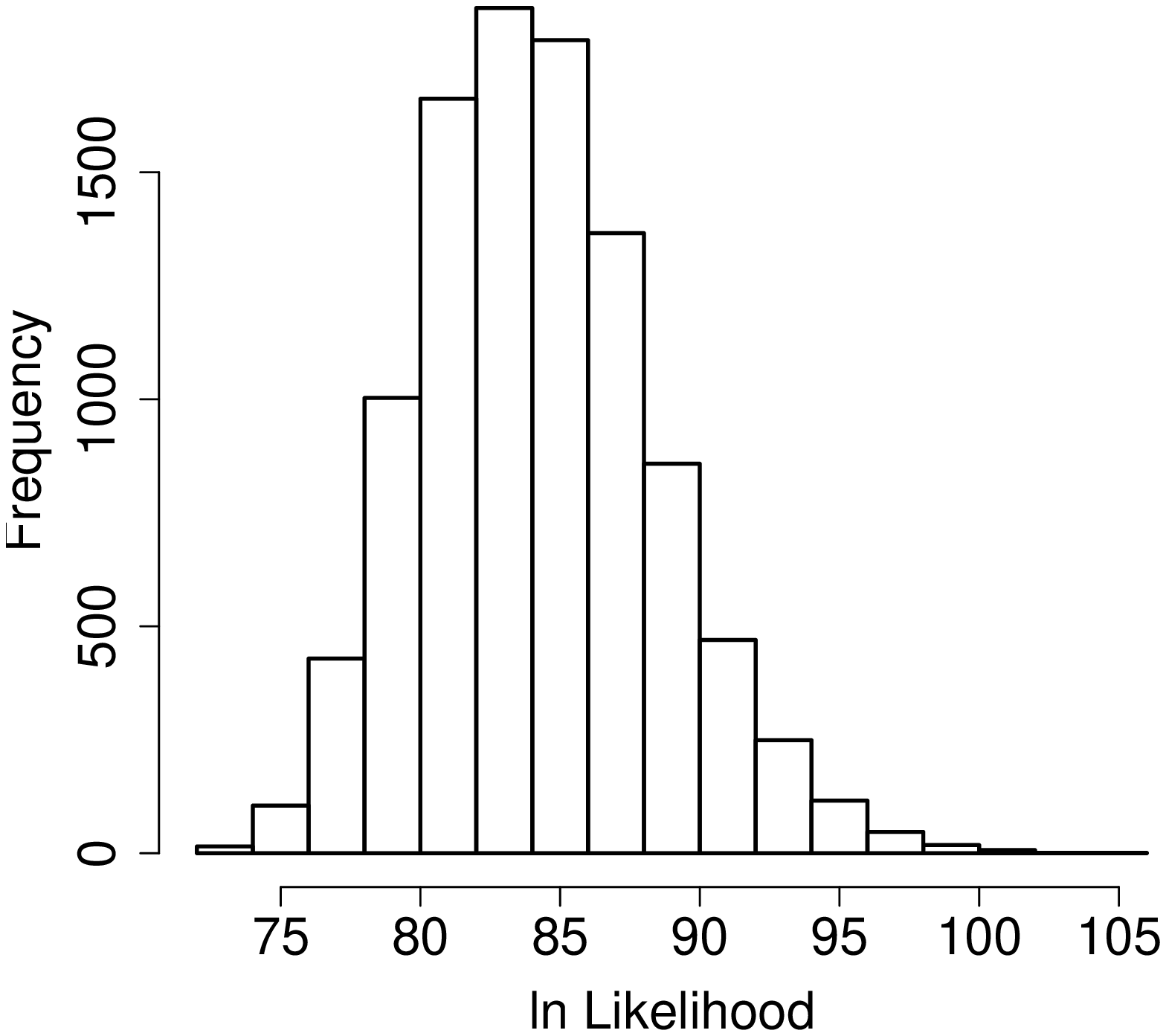}}
\caption{Fig.~\newFigure{f:Lemu}. Estimated distribution of the $\ln{\cal L}$ test 
statistic for the null hypothesis in the example of Fig.~\refno{example1}.} 
}

There are many other possible tests that could be considered, for example,
schemes that ``partition'' the $\chi^2$ to select sensitivity to different
characteristics~[\newCite{b:Best94}{D.~J.~Best, {\sl Nonparametric Comparison of Two Histograms}, Biometrics {\bf 50} (1994) 538.}].
   
\section{s:null}{Distributions Under the Null Hypothesis}

For the situation where the asymptotic distribution may not be good enough,
we would like to know the probability distribution of our test statistic
under the null hypothesis. However, we encounter a difficulty: our null
hypothesis is not completely specified! The problem is that the
distribution depends on the values of $\nu_i=a\mu_i$. Our null hypothesis
only says $\nu_i=a\mu_i$, but says nothing about what $\mu_i$ might be.
Note that it also doesn't specify $a$, but we have already discussed
that complication, which appears manageable (although in extreme
situations one might need to check for dependence on $a$).

We turn once again to the data to make an estimate for $\mu_i$, to be
used in estimating the distribution of our test statistics. The 
straightforward approach is to use the
maximum likelihood parameter estimators (under $H_0$):
$$\eqalign{
 \hat\mu_i &= {1\over 1+\hat a}(u_i+v_i),\cr
 \hat\nu_i &= {\hat a \over 1 + \hat a }(u_i+v_i),\cr
 }$$
where $\hat a = N_v/N_u$. The data is then repeatedly simulated using these
values for the parameters of the sampling distribution. For each simulation,
a value of the test statistic is obtained. The distribution so obtained is then
an estimate of the distribution of the test statistic under the null 
hypothesis, and $P$-values may be computed from this. Variations in the
estimates for $\hat\mu_i$ and $\hat a$ may be used to check robustness
of the probability estimates obtained in this way.

We have just described the approach that was used to compute the
estimated probabilities for the example of Fig.~\refno{example1}.
The bin contents in this case are reasonably large, and this 
approach works well enough for this case.

Unfortunately, this approach does very poorly in the low-statistics
realm. We consider a simple test case: Suppose our data is sampled
from a flat distribution with a mean of 1 count in each of 100 bins.
We test how well our estimated null hypothesis works for any given test statistic, $T$,
as follows:
\item{1.} Generate a pair of histograms according to the distribution just described.
\itemitem{(a)} Compute $T$ for this pair of histograms.
\itemitem{(b)} Given the pair of histograms, compute the estimated null hypothesis
according to the spcified prescription above.
\itemitem{(c)} Generate many pairs of histograms according to the estimated null hypothesis
in order to obtain an estimated distribution for $T$.
\itemitem{(d)} Using the estimated distribution for $T$, determine the estimated $P$-value
for the value of $T$ found in step 1a.
\item{2.} Repeat step 1 many times and make a histogram of the estimated $P$-values.
Note that this histogram should be uniform if the estimated $P$-values are
good estimates.

\vbox{
The distributions of the estimated probabilities for the seven test statistics under the null
hypothesis are shown in the second column of Fig.~\refno{f:H0toyAll}. If the null hypothesis
were to be rejected at the estimated 0.01 probability, this algortihm would actually reject $H_0$
19\% of the time for the $\chi^2$ statistic, 16\% of the time for the BDM
statistic, 24\% of the time for the $\ln\lambda$ statistic, and 29\% of the time 
for the ${\cal L}$ statistics, all
unacceptably larger than the desired 1\%. The KS, CVM, and AD statistics are 
all consistent with the desired 1\%. For comparison, the first column of Fig.~\refno{f:H0toyAll}
shows the distribution for a ``large statistics'' case, where sampling is from 
histograms with a mean of 100 counts in each bin. We find that all test statistics
display the desired flat distribution in this case. 
Table~\refno{t:nullHypScaleDep}\ summarizes these results.

\bigskip

\vbox{
\caption{Table~\newTable{t:nullHypScaleDep}. Probability that
the null hypothesis will be rejected with a cut at 1\% on the 
estimated distribution (see text). $H_0$ is estimated with the
bin-by-bin algorithm in the first two columns, by the uniform
histogram algorithm in the third column, and with a Gaussian kernel
estimation in the fourth column.}
\hbox{\centerline{\vbox{ 
\halign{#\hfil &\quad # & # & # & #\hfil\cr
\noalign{\vskip3pt\hrule\vskip3pt}
Test statistic & Probability (\%)& Probability (\%)& Probability (\%)& Probability (\%)\cr
\noalign{\vskip3pt\hrule\vskip4pt}
Bin mean = & 100             & 1 & \hfil 1 (uniform)\hfil & \hfil 1 (kernel)\hfil\cr
$H_0$ estimate & bin-by-bin & bin-by-bin & uniform & kernel\cr 
\noalign{\vskip3pt\hrule\vskip3pt}
$\chi^2$   & $0.97\pm0.24$ & $18.5\pm1.0$ & $1.2\pm0.3$ &  $1.33\pm0.28$\cr
BDM        & $0.91\pm0.23$ & $16.4\pm0.9$ & $0.30\pm0.14$ & $0.79\pm0.22$ \cr
KS         & $1.12\pm0.26$ & $0.97\pm0.24$ & $1.0\pm0.2$ & $1.21\pm0.27$\cr
CVM        & $1.09\pm0.26$ & $0.85\pm0.23$ & $0.8\pm0.2$ & $1.27\pm0.28$\cr
AD         & $1.15\pm0.26$ & $0.85\pm0.23$ & $1.0\pm0.2$ & $1.39\pm0.29$\cr
$\ln\lambda$ & $0.97\pm0.24$ & $24.2\pm1.1$ & $1.5\pm0.3$ & $2.0\pm0.34$\cr
$\ln{\cal L}$ & $0.97\pm0.24$ & $28.5\pm1.1$ & $0.0\pm0.0$ & $0.061\pm0.061$\cr
\noalign{\vskip3pt\hrule\vskip3pt}
}}}}}
}

\vfil\break

\vbox{
\centerline{\includegraphics[width=6in]{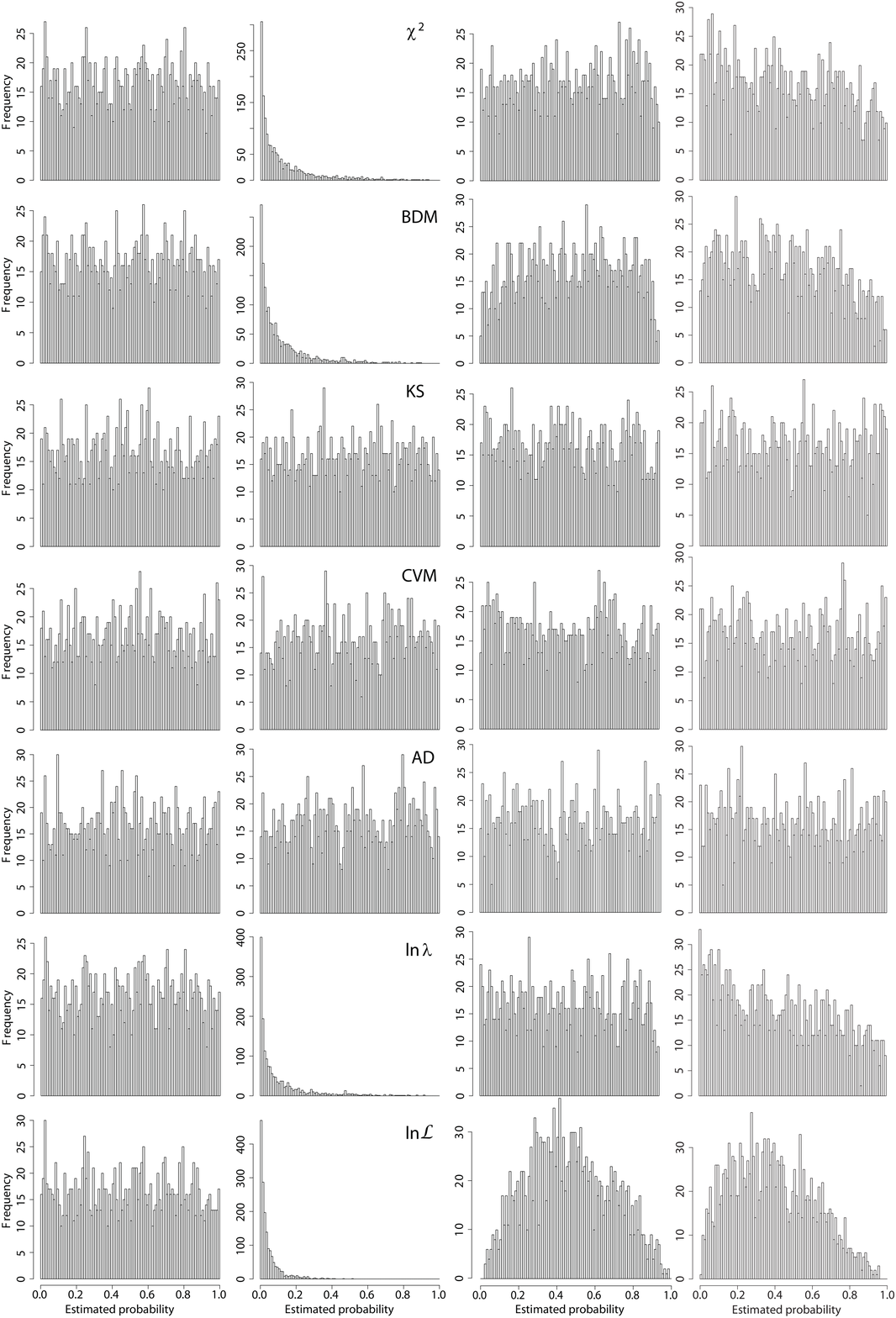}}
\caption{Fig.~\newFigure{f:H0toyAll}. See caption on next page}
}

\caption{Fig.~\refno{f:H0toyAll}. (Figure on previous page) Distribution of the 
estimated probability that the test statistic is worse
than that observed, for seven different test statistics. The data are generated according
to the null hypothesis, consisting of 100 bin histograms with a mean of 100 counts (left 
column) or one count (other columns). The first and second columns are for an estimated H0
computed as the weighted bin-by-bin average. The third column is for an estimated H0 where
each bin is the average of the total contents of both histograms, divided by the
number of bins. The rightmost column is for an estimated H0 estimated with
a Gaussian kernel estimator using the
contents of both histograms. The $\chi^2$ is computed without
combining bins.}

It may be noted that the issue really is one appearing at low statistics. 
We can give some intuition for the observed effect. Consider the likely scenario
at low statistics that some bins will have zero counts in both histograms.
In this case our algorithm for the estimated null hypothesis yields a zero
mean for these bins. The simulation used to determine the
probability distribution for the test statistic will always have zero counts in 
these bins, that is, there will always be agreement between the two 
histograms in these bins. Thus, the simulation will find that values of the test statistic
are more probable than it should. 

If we
tried the same study with, say, a mean of 100 counts per bin, we would find
that the probability estimates are valid, at least this far into the
tails.  The left column of Fig.~\refno{f:H0toyAll} shows that more sensible behavior is achieved with
larger statistics.
The $\chi^2$, $\ln\lambda$, and $\ln{\cal L}$ statistics perform essentially identically at high statistics, as expected, since in the normal approximation they are equivalent.  

The AD, CVM, and KS tests are more robust under our estimates of $H_0$ than the others, as they tend to emphasize
the largest differences and are not so sensitive to bins that always agree.
For these statistics, we see that our procedure for estimating
$H_0$ does well even for low statistics, although we caution again that we are not
examining the far tails of the distribution.

There are various possible approaches to salvaging the situation
in the low statistics regime. Perhaps the simplest is to rely on the
typically valid assumption that the underlying $H_0$ distribution is
``smooth''. Then instead of having an unknown parameter for each
bin, we only need to estimate a few parameters to describe the
smooth distribution, and effectively more statistics are available.
 
For example, we may repeat the algorithm for our example of a
mean of one count per bin,
but now assuming a smooth background represented by a uniform
distribution. This is cheating a bit, since we perhaps aren't supposed to
know that this is really what we are sampling from, but we'll
pretend that we looked at the data and decided that this was plausible.
As usual, we would in practice want to try other possibilities to
evaluate systematic effects. 

Thus, we estimate:
$$\eqalign{
 \hat\mu_i &= N_u/k,\ i=1,2,\ldots,k\cr
 \hat\nu_i &= N_v/k,\ i=1,2,\ldots,k.\cr
 }$$
The resulting distributions for the estimated probabilities
are shown in the third column of Fig.~\refno{f:H0toyAll}.
These distributions are much more reasonable, at least at the 
level of a per cent (1650 sample experiments are generated in each case,
and the estimated $P$ value is estimated for each experiment with 
1650 evaluations of the null hypothesis for that experiment). 

It should be remarked that the $\ln{\cal L}$ and, perhaps, to a much lesser
extent the BDM statistic, do not give the desired 1\% result, but now
err on the ``conservative'' side. It may be possible to mitigate this with a different 
algorithm, but this has not been investigated. We may expect the power of these
statistics to suffer under the approach taken here.

Since we aren't supposed to know that our null distribution is uniform, 
we also try another approach to get a feeling for whether we can really
do a legitimate analysis. Thus, we try a kernel estimator for the null distribution,
using the sum of the observed histograms as input. In this case, we have chosen
a Gaussian kernel, with a standard deviation of 2. The ``density'' package in
R [\refno{R}] is used for this. An example of such a kernel estimated
distribution is shown in Fig.~\refno{f:H0GKerneldensBW2}.
The resulting estimated probability distributions
of our test statistics are shown in the rightmost column of Fig.~\refno{f:H0toyAll}.
In general, this works pretty well. The bandwidth was chosen here to be rather small; 
a larger bandwidth would presumably improve the results. 

\vbox{
\centerline{\includegraphics[width=3.5in]{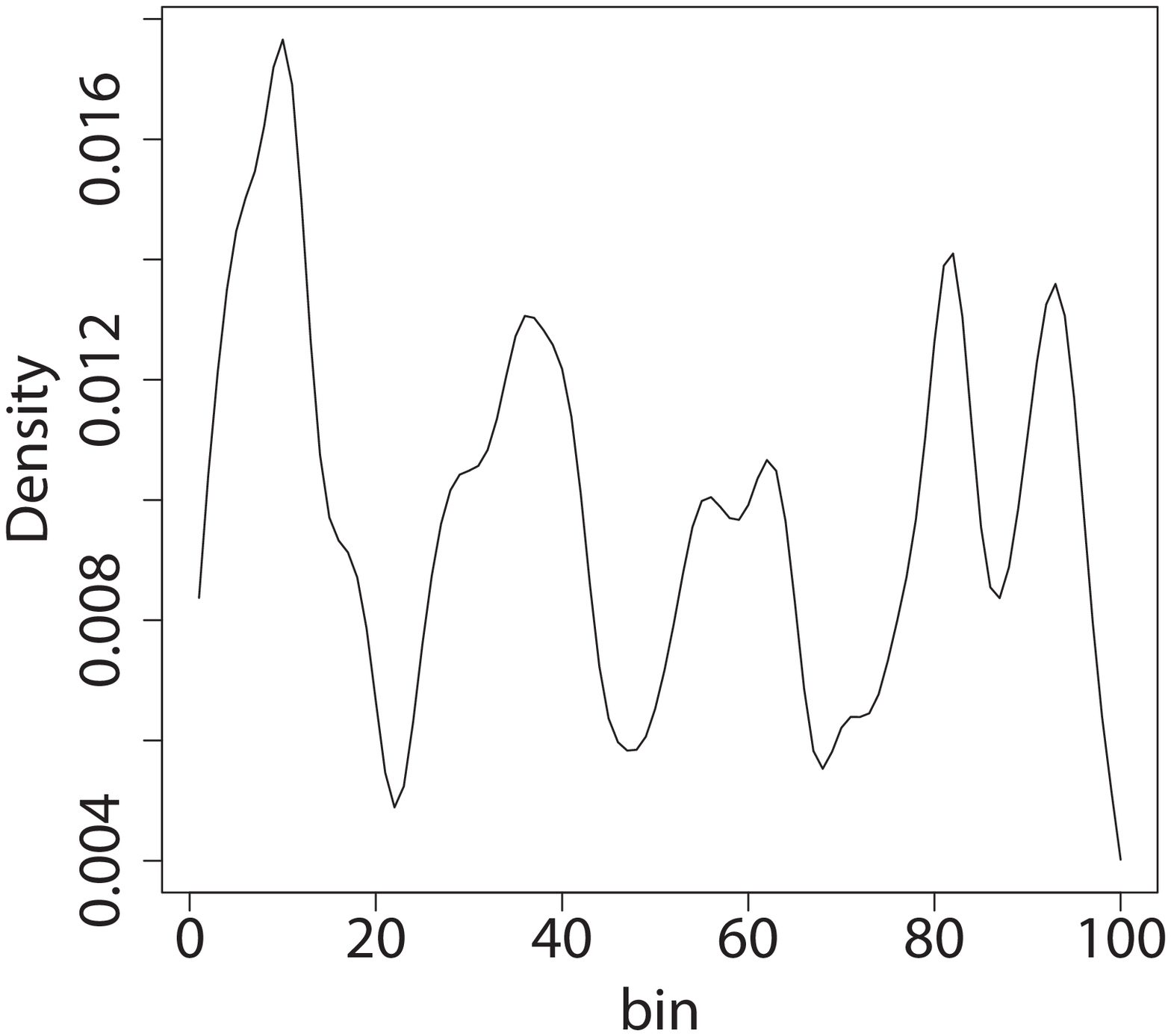}}
\caption{Fig.~\newFigure{f:H0GKerneldensBW2}. Sample Gaussian kernel density
estimate of the null hypothesis (for sampling from a true null).}
}

\bigskip

\section{s:power}{Comparison of Power of Tests}

The power depends on what the alternative hypothesis is. Here, we
mostly investigate adding a Gaussian component on top of a uniform background
distribution. This choice is motivated by the scenario where one distribution
appears to show some peaking structure, while the other does not. We also
look briefly at a different extreme, that of a rapidly varying alternative.

The data for this study are generated as follows:
The background (null distribution) has a mean of one event per histogram bin.
The Gaussian has a mean of 50 and a standard deviation of 5, in units of
bin number. We vary the amplitude of the Gaussian and count how often
the null hypothesis is rejected at the 1\% confidence level. The amplitude
is measured in percent, for example a 25\% Gaussian has a total amplitude 
corresponding to an  
average of 25\% of the total counts in the histogram, including the
(small) tails extending beyond the histogram boundaries. The Gaussian
counts are added to the counts from the null distribution. An example is
shown in Fig.~\refno{f:H0H1for25on1}.

\vbox{
\includegraphics[width=2.5in]{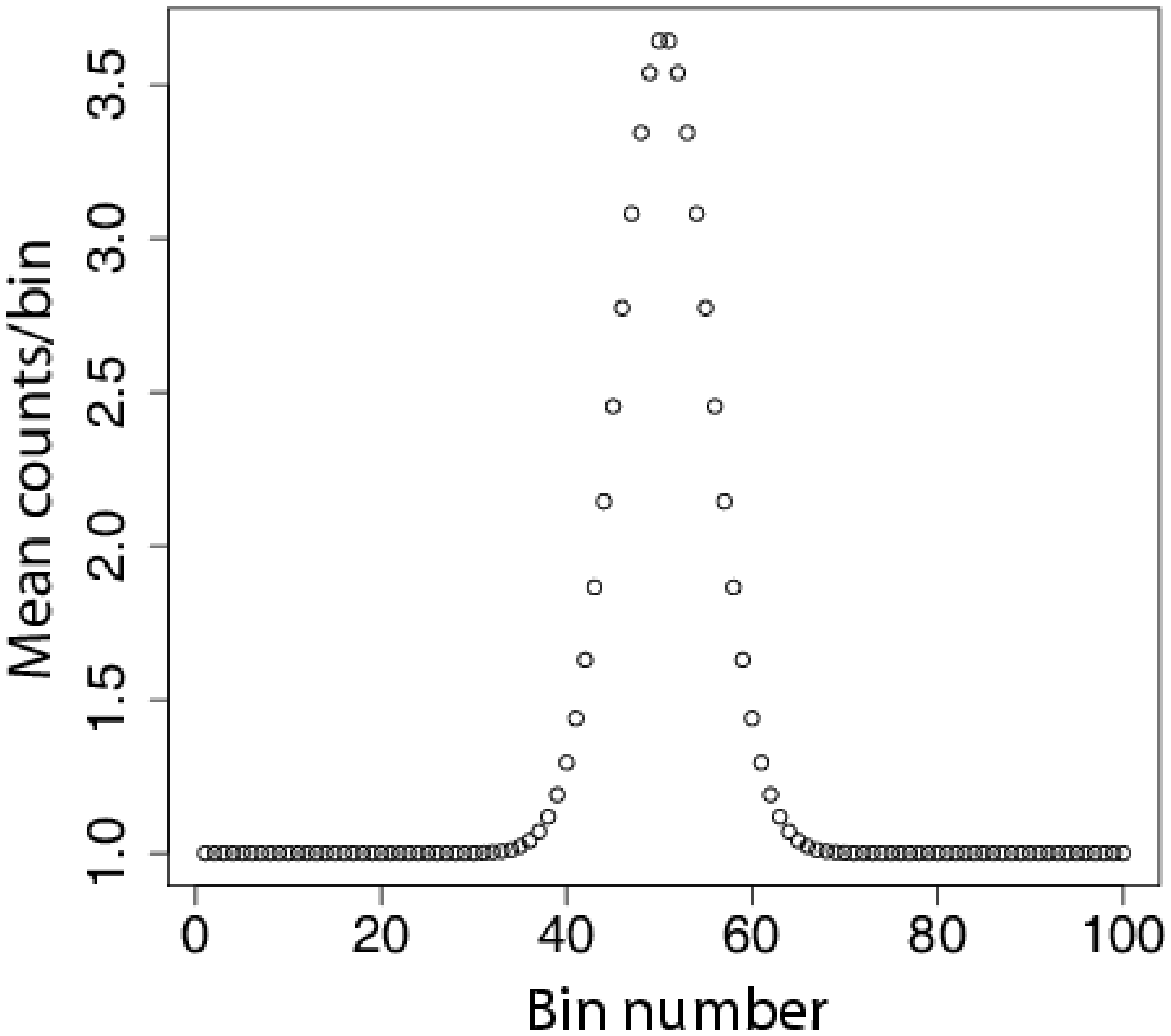}
\includegraphics[width=2.5in]{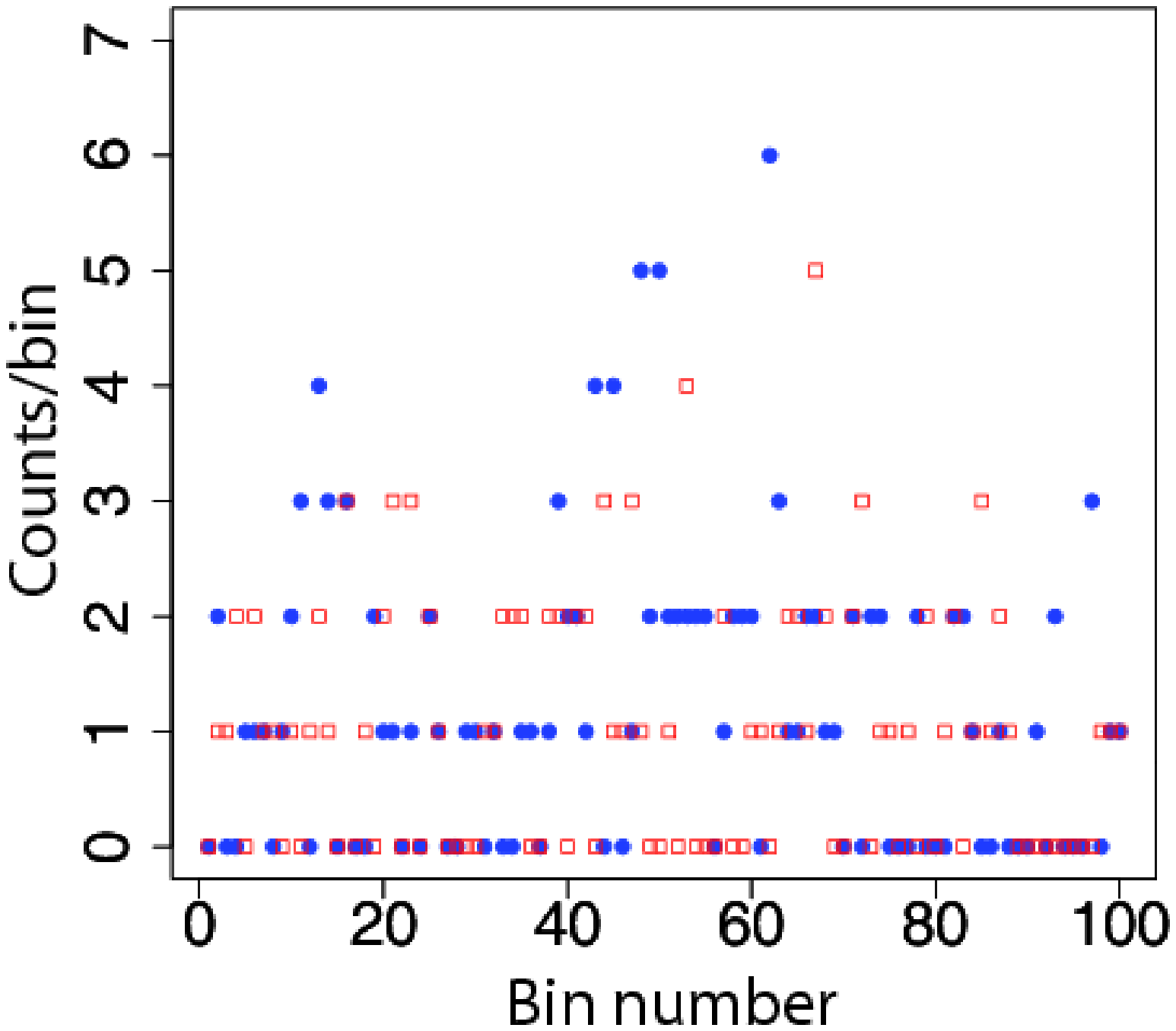}
\caption{Fig.~\newFigure{f:H0H1for25on1}. Left: The mean bin 
contents for a 25\% Gaussian on a flat background of one count/bin (note the suppressed zero).
Right: Example sampling from the 25\% Gaussian (filled blue dots) and
from the uniform background (open red squares).}
 
The distribution of estimated probability, under $H0$, that the test statistic is worse
than that observed (i.e., the distribution of $P$-values) 
is shown in Fig.~\refno{f:power} for seven different test statistics.
Three different magnitudes of the Gaussian amplitude are displayed.
The power of the tests to reject the null hypothesis at the 99\% confidence level is
summarized in Table~\refno{t:power}\ and in Fig.~\refno{f:powerSum}\ for 
several different alternative hypothesis amplitudes. 

\bigskip

\caption{Table~\newTable{t:power}. Estimates of power for seven different 
test statistics, as a function of $H_1$. The comparison histogram ($H_0$) is
generated with all $k=100$ bins Poisson of mean 1. The selection is at the 99\% 
confidence level, that is, the null hypothesis is accepted with (an estimated) 99\% probability
if it is true.}
\centerline{\vbox{ 
  \hrule\vskip2pt
\halign{#\hfil\quad&\hfil#\hfil&\hfil#\hfil&\hfil#\hfil&\hfil#\hfil&\hfil#\hfil&\hfil#\hfil\cr
          & H0            & 12.5        & 25           & 37.5         & 50          & -25\cr  
\noalign{\vskip-2pt}
Statistic  &\% &\% &\% &\% &\% &\% \cr 
  \noalign{\vskip2pt\hrule\vskip2pt}
$\chi^2$            & $1.2\pm0.3$   & $1.3\pm0.3$ & $4.3\pm0.5$  & $12.2\pm0.8$ & $34.2\pm1.2$ & $1.6\pm0.3$ \cr
BDM   & $0.30\pm0.14$ & $0.5\pm0.2$ & $2.3\pm0.4$  & $10.7\pm0.8$ & $40.5\pm1.2$ & $0.9\pm0.2$\cr
KS                  & $1.0\pm0.2$   & $3.6\pm0.5$ & $13.5\pm0.8$ & $48.3\pm1.2$ & $91.9\pm0.7$ & $7.2\pm0.6$\cr
CVM                 & $0.8\pm0.2$   & $1.7\pm0.3$ & $4.8\pm0.5$ & $35.2\pm1.2$ & $90.9\pm0.7$ & $2.7\pm0.4$\cr
AD                  & $1.0\pm0.2$   & $1.8\pm0.3$ & $6.5\pm0.6$ & $42.1\pm1.2$ & $94.7\pm0.6$ & $2.8\pm0.4$\cr
$\ln\lambda$        & $1.5\pm0.3$   & $1.9\pm0.3$ & $6.4\pm0.6$  & $22.9\pm1.0$ & $67.1\pm1.2$ & $2.4\pm0.4$ \cr
$\ln{\cal L}$       & $0.0\pm0.0$   & $0.1\pm0.1$ & $0.8\pm0.2$ & $6.5\pm0.6$  & $34.8\pm1.2$ & $0.0\pm0.0$\cr
}
  \vskip2pt\hrule
}}
}

\vfil\break

\vbox{
\centerline{\includegraphics[width=4.4in]{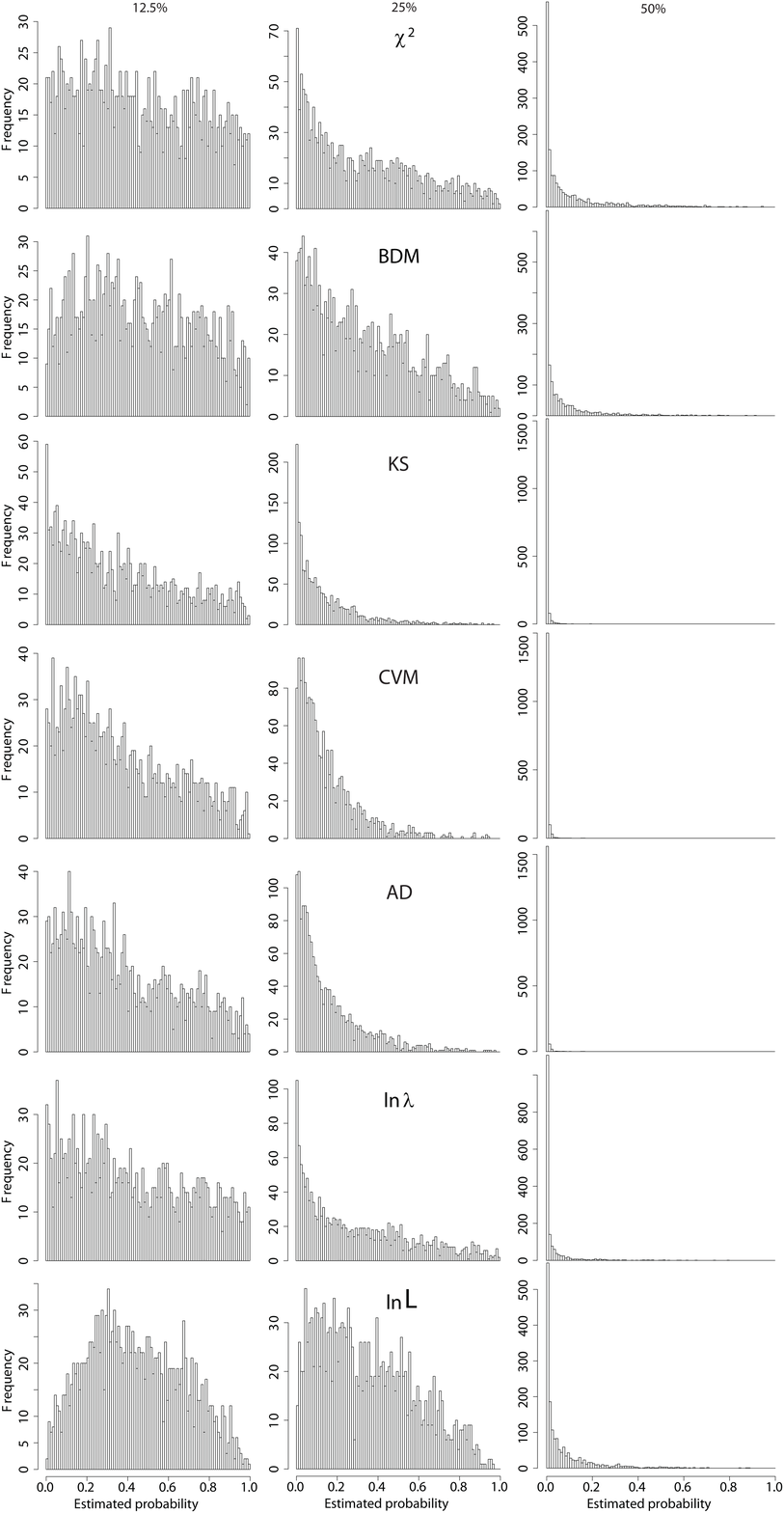}}
\caption{Fig.~\newFigure{f:power}. See caption, next page.}
}

\caption{Fig.~\refno{f:power}. (Figure on previous page) Distribution of estimated probability, under $H0$, that the test statistic is worse
than that observed, for seven different test statistics. The data are generated according
to a uniform distribution, consisting of 100 bin histograms with a mean of 1 count, for one
histogram, and for the other histogram with a uniform distribution plus a
Gaussian of strength 12.5\% (left 
column), 25\% (middle column), and 50\% (right column). The $\chi^2$ is computed without
combining bins.}

\bigskip

\vbox{
\centerline{\includegraphics[width=6.2in]{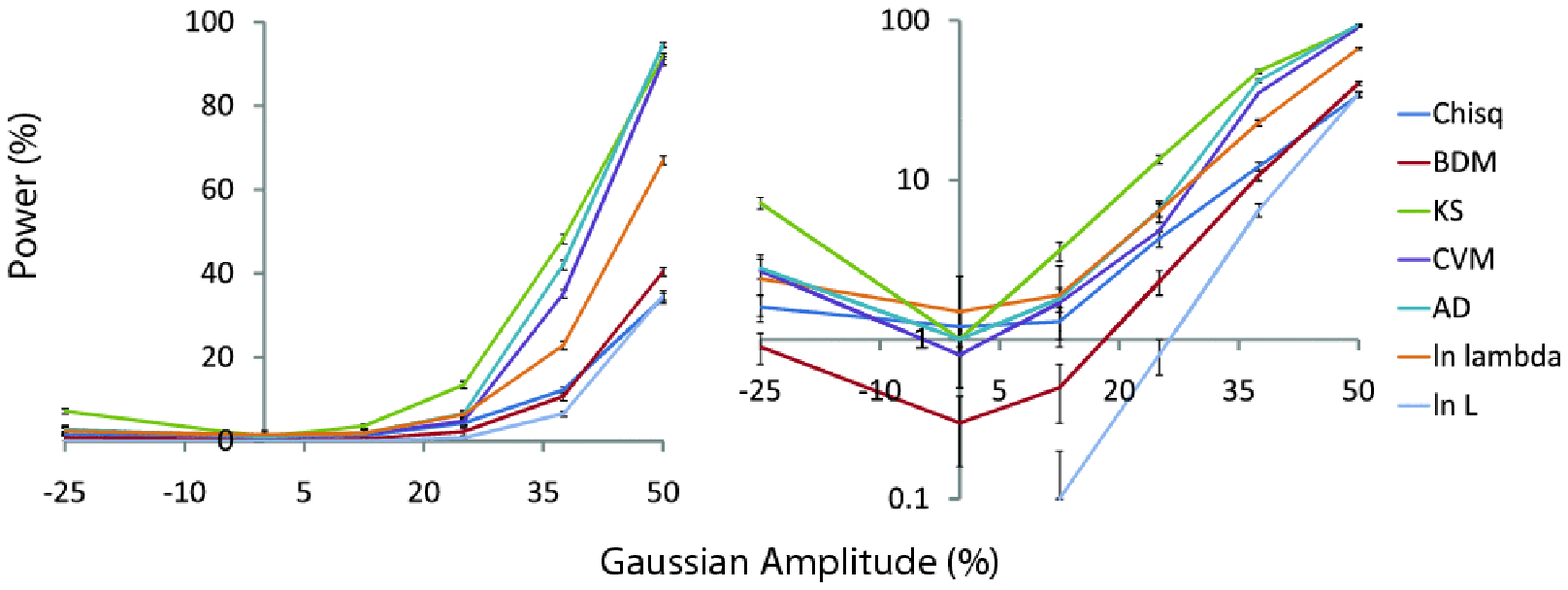}}
\caption{Fig.~\newFigure{f:powerSum}. Summary of power of seven different test
statistics, for the alternative hypothesis with a Gaussian bump. Left: linear vertical scale; 
Right: logarithmic vertical scale. [Best viewed in color. At an amplitude of 35\%,
the ordering, from top to bottom, of the curves is: KS, AD, CVM, $\ln\lambda$,
$\chi^2$, BDM, $\ln{\cal L}$.]}
}

\bigskip
 
In Table~\refno{t:power100}\ we take a look at the performance of our seven
statistics for histograms with large bin contents.
It is interesting that in this large-statistics case, for the $\chi^2$ and similar tests, the power to 
reject a dip is greater than the power to reject a bump of the same area.
This is presumably because the ``error estimates'' for the $\chi^2$
are based on the square root of the observed counts, and hence give
smaller errors for smaller bin contents.  We also observe that the comparative
strength of the KS, CVM, and AD tests versus the $\chi^2$, BDM, $\ln\lambda$, and
$\ln{\cal L}$ tests in the small statistics situation is largely reversed
in the large statistics case. 

\caption{Table~\newTable{t:power100}. Estimates of power for seven different 
test statistics, as a function of $H_1$. The comparison histogram ($H_0$) is
generated with all $k=100$ bins Poisson of mean 100. The selection is at the 99\%
confidence level.}
\centerline{\vbox{ 
  \hrule\vskip2pt
\halign{#\hfil\quad&\hfil#\hfil&\hfil#\hfil&\hfil#\hfil\cr
           & H0             & 5            & -5\cr  
\noalign{\vskip-2pt}
Statistic &\% &\% &\% \cr 
  \noalign{\vskip2pt\hrule\vskip2pt}
$\chi^2$            & $0.91\pm0.23$  & $79.9\pm1.0$ & $92.1\pm0.7$ \cr
BDM                 & $0.97\pm0.24$  & $80.1\pm1.0$ & $92.2\pm0.7$\cr
KS                  & $1.03\pm0.25$  & $77.3\pm1.0$ & $77.6\pm1.0$\cr
CVM                 & $0.91\pm0.23$  & $69.0\pm1.1$ & $62.4\pm1.2$\cr
AD                  & $0.91\pm0.23$  & $67.5\pm1.2$ & $57.8\pm1.2$\cr
$\ln\lambda$        & $0.91\pm0.23$  & $79.9\pm1.0$ & $92.1\pm0.7$\cr
$\ln{\cal L}$       & $0.97\pm0.24$  & $79.9\pm1.0$ & $91.9\pm0.7$\cr
}
  \vskip2pt\hrule
}}

To get an idea of what happens for a radically different alternative to the null
distribution, we consider sensitivity to sampling from the ``sawtooth'' distribution as shown in
figure~\refno{f:H0H1for50on1saw}.  This is to be compared once again to samplings
from the uniform histogram. The results are tabulated in Table~\refno{t:powersaw}.
The ``percentage'' sawtooth here refers to the fraction of the
null hypothesis mean. That is, a 100\% sawtooth on a 1 count/bin background oscillates between
a mean of 0 counts/bin and 2 counts/bin. The period of the sawtooth is always 
two bins.

\includegraphics[width=2.5in]{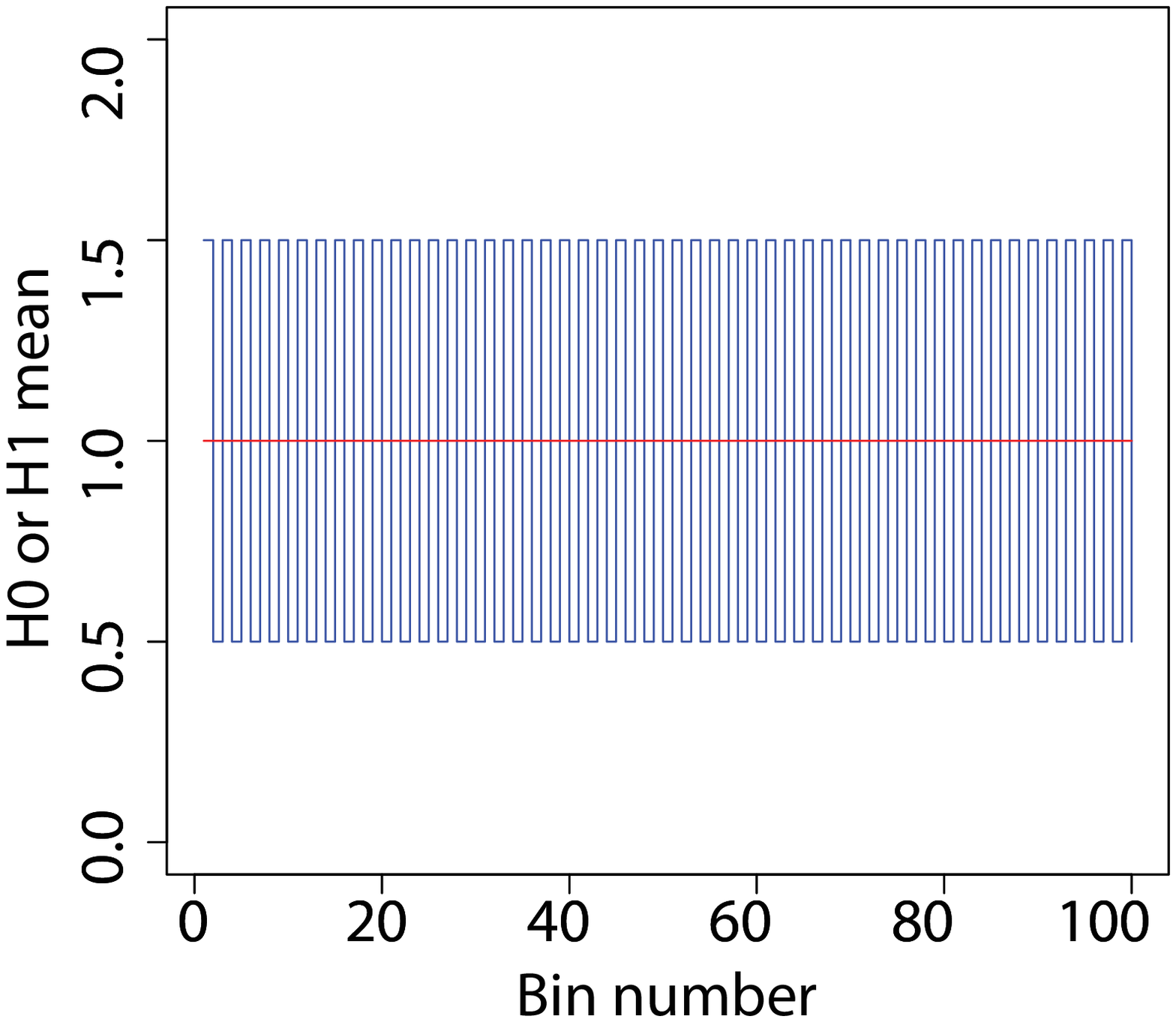}
\includegraphics[width=2.5in]{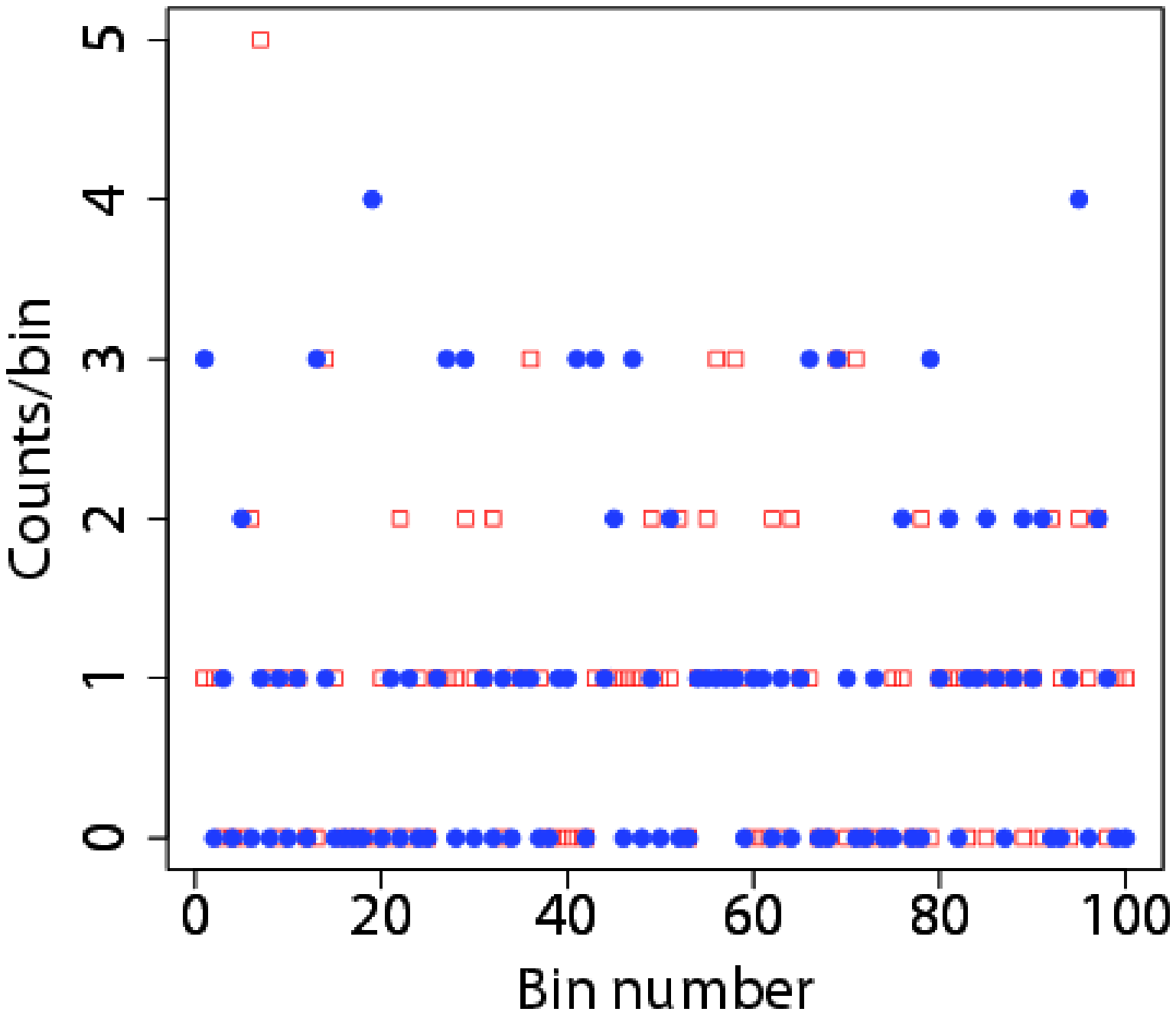}
\caption{Fig.~\newFigure{f:H0H1for50on1saw}. Left: The mean bin 
contents for a 50\% sawtooth on a flat background of one count/bin (blue),
compared with the flat background means (red).
Right: Example sampling from the 50\% sawtooth (filled blue dots) and
from the uniform background (open red squares).} 

\vbox{ 
In this example, the $\chi^2$ and likelihood ratio tests are the clear winners, with BDM next. The KS, CVM, and AD tests reject the null hypothesis with the same probability as for sampling from a true null distribution. This very poor performance for these tests
is readily understood, as these tests are all based on the cumulative distributions, which average out local oscillations.

\bigskip

\caption{Table~\newTable{t:powersaw}. Estimates of power for seven different 
test statistics, for a ``sawtooth'' alternative distribution.}
\centerline{\vbox{ 
  \hrule\vskip2pt
\halign{#\hfil\quad&\hfil#\hfil&\hfil#\hfil&\hfil#\hfil&\hfil#\hfil&\hfil#\hfil\cr
         &  50 & 100\cr  
\noalign{\vskip-2pt}
 Statistic &\% &\%  \cr 
  \noalign{\vskip2pt\hrule\vskip2pt}
$\chi^2$            & $3.7\pm0.5$   & $47.8\pm1.2$ \cr
BDM   & $1.9\pm0.3$   & $33.6\pm1.2$ \cr
KS                  & $0.85\pm0.23$ & $1.0\pm0.2$  \cr
CVM                 & $0.91\pm0.23$ & $1.0\pm0.2$  \cr
AD                  & $0.91\pm0.23$ & $1.2\pm0.3$  \cr
$\ln\lambda$        & $4.5\pm0.5$   & $49.6\pm1.2$ \cr
$\ln{\cal L}$       & $0.30\pm0.14$ & $10.0\pm0.7$  \cr
}
  \vskip2pt\hrule
}}
}

\bigskip

\chapter{ch:conclusions}{Conclusions}

These studies have demonstrated some important lessons in ``goodness-of-fit'' testing:
 
\item{1.} There is no single ``best'' test for all applications. 
Statements such as ``test X is better than test Y'' are empty without giving more context. For example,
the Anderson-Darling test is often very powerful in testing normality of data against alternatives with non-normal
tails (such as the Cauchy distribution) [\newCite{b:Stephens74}{M.~A.~Stephens, {\sl EDF Statistics for Goodness of Fit and Some Comparisons}, Jour. Amer. Stat. Assoc. {\bf69} (1974) 730.}]. However, we have seen
that it is not always especially powerful in other situations. The more we know about
what we wish to test for, the more reliably we can choose a powerful test. Each of the tests investigated
here may be reasonable to use, depending on the circumstance. Even the controversial ${\cal L}$ test works as well
as the others sometimes. However, there is no known situation where it actually performs better than all of the others, and 
indeed the situations where it is observed to perform as well are here limited to those where it is equivalent to
another test.

\item{2.} Computing probabilities via simulations is a very useful technique. However, it must be done with care.
The issue of tests with an incompletely specified null hypothesis is particularly insidious. Simply generating
a distribution according to some assumed null distribution can lead to badly wrong results. Where this could occur,
it is important to verify the validity of the procedure. Note that we have only looked into the tails to the
1\% level. The validity must be checked to whatever level of probability is needed for the results. Thus, we
cannot blindly assume the results quoted here at the 1\% level will still be true at, say, the 0.1\% level.

We have concentrated in this paper on the specific question of comparing two histograms. However, the general
considerations apply more generally, to testing whether two datasets are consistent with being drawn from
the same distribution, and to testing whether a dataset is consistent with a predicted distribution. The
KS, CVM, AD, $\ln{\cal L}$, and ${\cal L}$ tests may all be constructed for these other situations (as well as the
$\chi^2$ and BDM, if we bin the data). 

\vskip1cm

\noindent {\bf References}

\makebib

\byebye